\documentclass[5p]{elsarticle}

\usepackage[flushleft]{threeparttable}
\usepackage[titlenumbered,linesnumbered,ruled]{algorithm2e}
\usepackage{lineno}
\usepackage{hyperref}
\usepackage{color}
\usepackage{xcolor}
\usepackage{booktabs}
\usepackage{longtable}
\usepackage{lipsum}
\usepackage{supertabular}
\usepackage{amssymb}
\usepackage{graphicx}
\usepackage{subfigure}
\usepackage{caption}
\usepackage{url}
\usepackage{mathtools}
\usepackage{multirow}
\usepackage{array}
\usepackage{footnote}
\usepackage{listings}
\usepackage{adjustbox}
\usepackage{lipsum}
\usepackage{float}

\usepackage{pifont}
\newcommand{\cmark}{\ding{51}}%
\newcommand{\xmark}{\ding{55}}%

\makesavenoteenv{tabular}
\makesavenoteenv{table}

\graphicspath{{figs/}}

\begin{document}

\begin{frontmatter}

\title{MalDy: Portable, Data-Driven Malware Detection using Natural Language 
    Processing and Machine Learning Techniques on Behavioral Analysis Reports} 
\author{ElMouatez Billah Karbab, Mourad Debbabi}
\address{Concordia University, Montreal, Canada}
\ead{e\_karbab@encs.concordia.ca}

\begin{abstract}
    In response to the volume and sophistication of malicious software or
    malware, security investigators rely on dynamic analysis for malware
    detection to thwart obfuscation and packing issues. Dynamic analysis is the
    process of executing binary samples to produce reports that summarise their
    runtime behaviors. The investigator uses these reports to detect malware
    and attribute threat type leveraging manually chosen features. However, the
    diversity of malware and the execution environments makes manual approaches
    not scalable because the investigator needs to manually engineer
    fingerprinting features for new environments. In this paper, we propose,
    MalDy (mal~die), a portable (plug and play) malware detection and family
    threat attribution framework using supervised machine learning techniques.
    The key idea of MalDy of portability is the modeling of the behavioral reports
    into a sequence of words, along with advanced natural language processing
    (NLP) and machine learning (ML) techniques for automatic engineering of
    relevant security features to detect and attribute malware without the
    investigator intervention. More precisely, we propose to use
    \textit{bag-of-words} (BoW) NLP model to formulate the behavioral reports.
    Afterward, we build ML ensembles on top of BoW features. We extensively
    evaluate MalDy on various datasets from different platforms (Android and
    Win32) and execution environments. The evaluation shows the effectiveness
    and the portability MalDy across the spectrum of the analyses and settings. 
\end{abstract}


\end{frontmatter}

\section{Introduction} 
\label{sec_intro}

Malware investigation is an important and time consuming task for security
investigators. The daily volume of malware raises the automation necessity of
the detection and threat attribution tasks. The diversity of platforms and
architectures makes the malware investigation more challenging. The
investigator has to deal with a variety of malware scenarios from Win32 to
Android. Also, nowadays malware targets all CPU architectures from x86 to ARM
and MIPS that heavily influence the binary structure. The diversity of malware
adopts the need for portable tools, methods, and techniques in the toolbox of
the security investigator for malware detection and threat attribution.

Binary code static analysis is a valuable tool to investigate malware in
general. It has been used effectively and efficiently in many solutions
\cite{arp2014drebin}, \cite{Hu13MutantX}, \cite{karbab2018maldozer},
\cite{DBLP:journals/corr/abs-1712-08996}, and \cite{Mariconti2017MaMaDroid} in
PC and Android realms. However, the use of static analysis could be problematic
on heavily obfuscated and custom packed malware. Solutions, such as
\cite{Hu13MutantX}, address those issues partially but they are
platform/architecture dependent and cover only simple evading techniques. On
the other hand, the dynamic analysis solutions \cite{Toward2017WillemsHF},
\cite{Bayer09Scalable}, \cite{Wong2016Intellidroid},
\cite{DynaLog2016Alzaylaee}, \cite{malrec2018Giorgio},
\cite{DBLP:journals/corr/abs-1806-08893}, \cite{DBLP:conf/IEEEares/KarbabD18},
\cite{Kernel2011Isohara}, \cite{Automatic2011Rieck},
\cite{Irolla2016Glassbox} show more robustness to evading techniques such as
obfuscation and packing. Dynamic analysis's main drawback is its hungry to
computation resources \cite{Wang2015DROIT}, \cite{Graziano2015Needles}. Also,
it may be blocked by anti-emulation techniques, but this is less common
compared to the binary obfuscation and packing. For this reason, dynamic (also
behavioral) analysis is still the default choice and the first analysis for
malware by security companies. 


The static and behavioral analyses are sources for security features, which the
security investigator uses to decide about the binary maliciousness. Manual
inspection of these features is a tedious task and could be automated using
machine learning techniques. For this reason, the majority of the
state-of-the-art malware detection solutions use machine learning
\cite{Mariconti2017MaMaDroid}, \cite{arp2014drebin}. We could classify these
solutions' methodologies into supervised and unsupervised. The supervised
approach, such as \cite{Nataraj2011comparative} for Win32 and
\cite{wu2014droiddolphin}, \cite{Sen2016StormDroid} for Android, is the most
used in malware investigation \cite{Martinelli2016find},
\cite{Sen2016StormDroid}. The supervised machine learning process starts by
training a classification model on a train-set. Afterward, we use this model on
new samples in a production environment. Second, the unsupervised approach,
such as \cite{Automatic2011Rieck}, \cite{karbab2016cypider},
\cite{Scalable2009Bayer}, \cite{DBLP:journals/corr/KarbabDAM17},
\cite{karbab2017dysign}, in which the authors cluster the malware samples into
groups based on their similarity. Unsupervised learning is more common in
malware family clustering \cite{Automatic2011Rieck}, and it is less common in
malware detection \cite{karbab2016cypider}.

In this paper, we focus on supervised machine learning techniques along with
behavioral (dynamic or runtime) analyses to investigate malicious software.
Dynamic and runtime analyses execute binary samples to collect their behavioral
reports. The dynamic analysis makes the execution in a sandbox environment
(emulation) where malware detection is an off-line task. The runtime analysis
is the process of collecting behavioral reports from production machines. The
security practitioner aims to obtain these reports to make an online malware
checking without disturbing the running system. 

\subsection{Problem Statement} 
\label{sec_problem_statement}

The state-of-the-art solutions, such as in \cite{Sen2016StormDroid},
\cite{kharraz2016unveil}, \cite{sgandurra2016automated}, rely on manual
security features investigation in the detection process. For example,
StormDroid \cite{Sen2016StormDroid} used \emph{Sendsms} and \emph{Recvnet}
dynamic features, which have been chosen base on a statistical analysis, for
Android malware detection. Another example, the authors in
\cite{Kolbitsch2009Effective} used explicit features to build behavior's graphs
for Win32 malware detection . The security features may change based on the
execution environment despite the targeted platform. For instance, the authors
\cite{Sen2016StormDroid} and \cite{DynaLog2016Alzaylaee} used different
security features due to the difference between the execution environments. In
the context of the security investigation, we are looking for a portable
framework for malware detection based on the behavioral reports across a
variety of platforms, architectures, and execution environments. The security
investigator would rely on this plug and play framework with a minimum effort.
We plug the behavioral analysis reports for the training and apply (play) the
produced classification model on new reports (same type) without an explicit
security feature engineering as in \cite{Sen2016StormDroid},
\cite{Kolbitsch2009Effective}, \cite{Chen2017Semi}; and this process works
virtually on any behavioral reports.  

\subsection{MalDy} 
\label{sec_solution}

We propose, MalDy, a portable and generic framework for malware detection and
family threat investigation based on behavioral reports. MalDy aims to be a
utility on the security investigator toolbox to leverage existing behavioral
reports to build a malware investigation tool without prior knowledge regarding
the behavior model, malware platform, architecture, or the execution
environment. More precisely, MalDy is portable because of the automatic mining
of relevant security features to allow moving MalDy to learn new environments'
behavioral reports without a security expert intervention.  Formally,
MalDy framework is built on top natural language processing (NLP) modeling and
supervised machine learning techniques. The main idea is to formalize the
behavioral report,  agnostically to the execution environment, into a bag of
words (BoW) where the features are the reports' words. Afterward, we leverage
machine learning techniques to automatically discover relevant security
features that help differentiate and attribute malware. The result is MalDy,
portable (Section \ref{sec_evaluation_portablity}), effective (Section
\ref{sec_evaluation_effectiveness}), and efficient (Section
\ref{sec_evaluation_efficiency}) framework for malware investigation.

\subsection{Result Summary} 
\label{sec_result_summary}

We extensively evaluate MalDy on different datasets, from various platforms,
under multiple settings to show the framework portability, effectiveness,
efficiency, and its suitability for general purpose malware investigation.
First, we experiment on Android malware behavioral reports of MalGenome
\cite{zhou2012dissecting}, Drebin \cite{arp2014drebin}, and Maldozer
\cite{karbab2018maldozer} datasets along with benign samples from AndroZoo
\cite{Allix2016AndroZoo} and PlayDrone
\footnote{https://archive.org/details/android\_apps} repositories. The reports
were generated using Droidbox \cite{droidbox_github} sandbox. MalDy achieved
99.61\%, 99.62\%, 93.39\% f1-score on the detection task on the previous
datasets respectively. Second, we apply MalDy on behavioral reports (20k
samples from 15 Win32 malware family) provided by ThreatTrack
security~\footnote{https://www.threattrack.com} (ThreatAnalyzer sandbox).
Again, MalDy shows high accuracy on the family attribution task, 94.86\%
f1-score, under different evaluation settings. Despite the difference between
the evaluation datasets, MalDy shows high effectiveness under the same
hyper-parameters with minimum overhead during the production, only 0.03 seconds
runtime per one behavioral report on modern machines. 

\subsection{Contributions}

\begin{itemize}

    \item \textbf{New Framework:} We propose and explore a data-driven approach
        to behavioral reports for malware investigation (Section
        \ref{sec_approach}). We leverage a word-based security feature
        engineering (Section \ref{sec_framework}) instead of the manual
        specific security features to achieve high portability across different
        malware platforms and settings.

    \item \textbf{BoW and ML}: We design and implement the proposed framework
        using the bag of words (BoW) model (Section
        \ref{sec_ensemble_composition}) and machine learning (ML) techniques
        (Section \ref{sec_ensemble_composition}). The design is inspired from
        NLP solutions where the word frequency is the key for feature
        engineering.

    \item \textbf{Application and Evaluation}: We utilize the proposed
        framework for Android Malware detection using behavioral reports from
        DroidBox \cite{droidbox_github} sandbox (Section \ref{sec_evaluation}).
        We extensively evaluate the framework on large reference datasets
        namely, Malgenome \cite{zhou2012dissecting},  Drebin
        \cite{arp2014drebin}, and Maldozer \cite{karbab2018maldozer} (Section
        \ref{sec_dataset}). To evaluate the portability, we conduct a further
        evaluation on Win32 Malware reports (Section
        \ref{sec_evaluation_portablity}) provided by a third-party security
        company. MalDy shows high accuracy in all the evaluation tasks.

\end{itemize}

\section{Threat Model} \label{sec_threat_model}

We position MalDy as a generic malware investigator tool. MalDy current design
considers only behavioral reports. Therefore, MalDy is by design resilient to
binary code static analysis issues like packing, compression, and dynamic
loading. MalDy performance depends on the quality of the collected reports. The
more security information and features are provided about the malware samples
in the reports the higher MalDy could differentiate malware from benign and
attribute to known families. The execution time and the random event generator
may have a considerable impact on MalDy because they affect the quality of the
behavioral reports. First, the execution time affect the amount of information
in the reports. Therefore, small execution time may result little information
to fingerprint malware. Second, random event generator may not produce the
right events to trigger certain malware behaviors; this loads to false
negatives. Anti-emulation techniques, used to evade Dynamic analysis,  
could be problematic for MalDy framework. However, this issue is related to the
choice the underlying execution environment. First, this problem is less
critical for a runtime execution environment because we collect the behavioral
reports from real machines (no emulation). This scenario presumes that all the
processes are benign and we check for malicious behaviors. Second, the security
practitioner could replace the sandbox tool with a resilient alternative since
MalDy is agnostic to the underlying execution environment.

\section{Overview} 
\label{sec_overview_intuition}

The execution of a binary sample (or app) produces textual logs whether in a
controlled environment (software sandbox) or production ones. The logs, a
sequence of statements, are the result of the app events, this depends on the
granularity of the logs. Furthermore, each statement is a sequence of words
that give a more granular description on the actual app event. From a security
investigation perspective, the app behaviors are summarized in an execution
report, which is a sequence of statements and each statement is a sequence of
words. We argue that malicious apps have distinguishable behaviors from benign
apps and this difference is translated into words in the behavioral report.
Also, we argue that similar malicious apps (same malware family) behaviors are
translated to similar words. 

\lstset{frame=tb,
  columns=flexible,
  basicstyle={\small\ttfamily},
  breaklines=true,
  tabsize=1
}
\begin{figure}[t]
\centering

\begin{lstlisting}
<open_key~key="HKEY_LOCAL_MACHINE\Software\Microsoft\Windows
NT\CurrentVersion\AppCompatFlags\Layers"/> <open_key
key="HKEY_CURRENT_USER\Software\Microsoft\Windows
NT\CurrentVersion\AppCompatFlags\Layers"/> <open_key
key="HKEY_LOCAL_MACHINE\System\CurrentControlSet\Services\
LanmanWorkstation\NetworkProvider"/>
</registry_section> <process_section> <enum_processes
apifunction="Process32First" quantity="84"/> <open_process targetpid="308"
desiredaccess="PROCESS_ALL_ACCESS PROCESS_CREATE_PROCESS PROCESS_CREATE_THREAD
PROCESS_DUP_HANDLE PROCESS_QUERY_INFORMATION PROCESS_SET_INFORMATION
PROCESS_TERMINATE PROCESS_VM_OPERATION PROCESS_VM_READ PROCESS_VM_WRITE
PROCESS_SET_SESSIONID PROCESS_SET_QUOTA SYNCHRONIZE"
apifunction="NtOpenProcess" successful="1"/>
\end{lstlisting}

\caption{Win32 Malware Behavioral Report Snippet 
                    (ThreatAnalyzer, www.threattrack.com)}
\label{fig_win32_malware_threat}
\end{figure}

Nowadays, there are many software sandbox solutions for malware investigations.
CWSandbox (2006-2011) is one of the first sandbox solutions for production use.
Later, CWSandbox becomes ThreatAnalyzer
\footnote{https://www.threattrack.com/malware-analysis.aspx}, owned by
ThreatTrack Security. TheatAnalyzer is a sandbox system for Win32 malware, and
it produces behavioral reports that cover most of the malware behavior aspects
such as a file, network, register access records. Figure
\ref{fig_win32_malware_threat} shows a snippet from a behavioral report
generated by ThreatAnalyzer. For android malware, we use \emph{DroidBox}
\cite{droidbox_github}, a well-established sandbox environment based on Android
software emulator \cite{android_emulator} provided by Google Android SDK
\cite{android_sdk}. Running an app may not lead to a sufficient coverage of the
executed app. As such, to simulate the user interaction with the apps, we
leverage \textit{MonkeyRunner} \cite{monkeyrunner}, which produces random UI
actions aiming for a broader execution coverage. However, this makes the app
execution non-deterministic since \textit{MonkeyRunner} generates random
actions. Figure \ref{fig_android_malware_droidbox} shows a snippet from the
behavioral report generated using DroidBox.

\lstset{frame=tb,
  columns=flexible,
  basicstyle={\small\ttfamily},
  breaklines=true,
  tabsize=1
}
\begin{figure}[t]
\centering

\begin{lstlisting}
"accessedfiles": { "1546331488": "/proc/1006/cmdline","2044518634":
"/data/com.macte.JigsawPuzzle.Romantic/shared_prefs/com.apperhand.global.xml",
"296117026":
"/data/com.macte.JigsawPuzzle.Romantic/shared_prefs/com.apperhand.global.xml",
"592194838": "/data/data/com.km.installer/shared_prefs/TimeInfo.xml",
"956474991": "/proc/992/cmdline"},"apkName": "fe3a6f2d4c","closenet":
{},"cryptousage": {},"dataleaks": {},"dexclass": { "0.2725639343261719": {
    "path": "/data/app/com.km.installer-1.apk", "type": "dexload"}
\end{lstlisting}

\caption{Android Malware Behavioral Report Snippet 
                (DroidBox \cite{droidbox_github})}
\label{fig_android_malware_droidbox}
\end{figure}

\begin{figure*}
   \centering
   \includegraphics[width=0.98\textwidth, trim=0cm 0cm 0cm 0cm, clip]
        {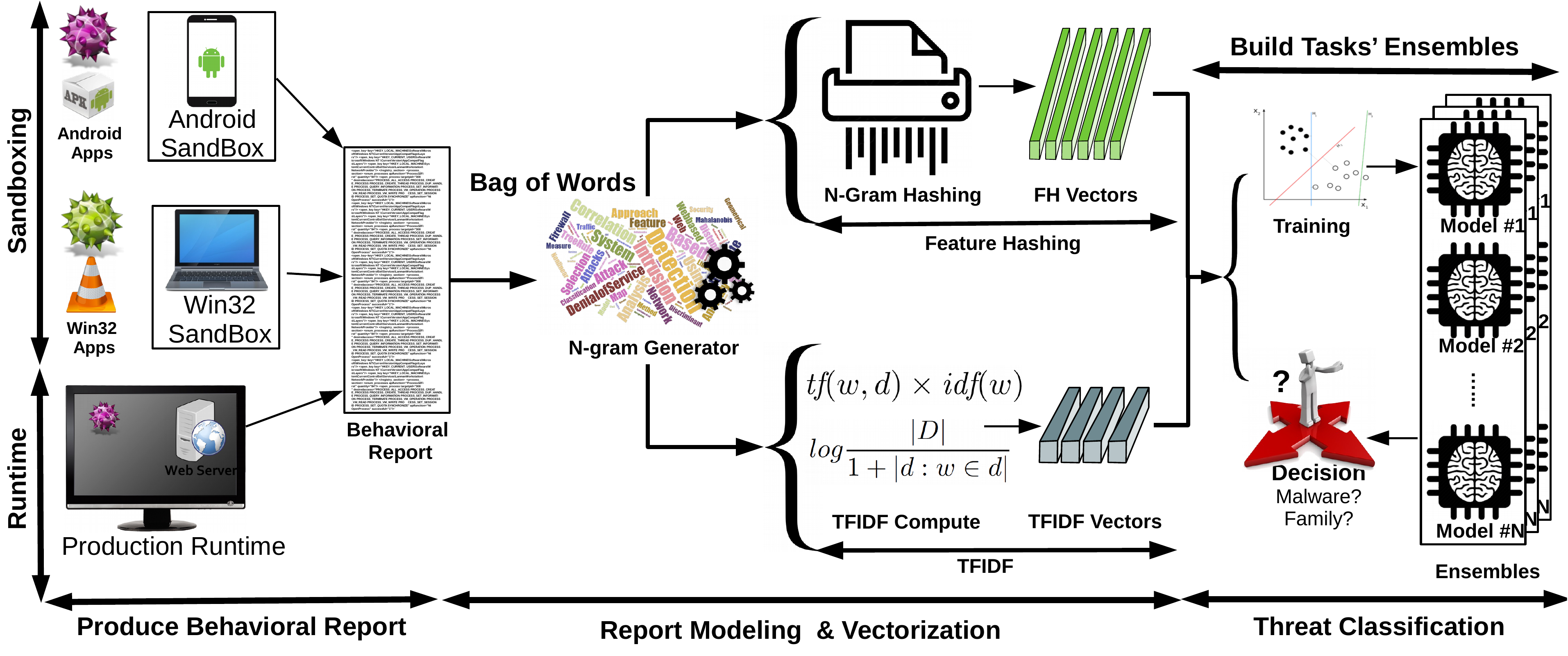}
   \caption{MalDy Methodology Overview}
   \label{fig_dysign_classification_approach_overview}
\end{figure*}

\section{Notation} 
\label{sec_notation}

\begin{itemize}
    \item $X = \{X_{build}, X_{test}\}:$ $X$ is the global dataset used to
        build and report MalDy performance in the various tasks. We use build
        set $X_{build}$ to train and tune the hyper-parameters of MalDy models.
        The test set $X_{test}$ is use to measure the final performance of
        MalDy, which is reported in the evaluation section. $X$ is divided
        randomly and equally to $X_{build}$ (50\%) and $X_{test}$ (50\%).
        To build the sub-datasets, we employ the stratified random split on the
        main dataset.

    \item $X_{build} = \{X_{train}, X_{valid}\}:$ Build set, $X_{build}$, is
        composed of the train set and validation set and used to build MalDy
        ensembles.

    \item $m_{build} = m_{train} + m_{valid}:$ Build size is the total number
        of reports used to build MalDy. The train set takes 90\% of the build
        set and the rest is used as a validation set . 

    \item $X_{train} = \{(x_0, y_0), (x_1, y_1), .., (x_{m_{train}},
        y_{m_{train}})\}:$ The train set, $X_{train}$, is the training dataset
        of MalDy machine learning models.

    \item $m_{train} = |X_{train}|:$ The size of $m_{train}$  is the number of
        reports in the train set.

    \item $X_{valid} = \{((x_0, y_0), (x_1, y_1), .., (x_{m_{valid}},
            y_{m_{valid}})\}:$ The validation set, $X_{valid}$, is the dataset
            used to tune the trained model. We choose the hyper-parameters that
            achieve the best scores on the validation set. 

    \item $m_{valid} = |X_{valid}|:$ The size of $m_{valid}$ is the number of
        reports in the validation set.

    \item $(x_i, y_i):$ A single record in $X$ is composed of a single report
        $x_i$ and its label $y_i \in \{+1, -1\}$. The label meaning depends on
        the investigation task. In the detection task, a positive means
        malware, and a negative means benign. In the family attribution task, a
        positive means the sample is part of the current model malware family
        and a negative is not.

    \item $X_{test} = \{((x_0, y_0), (x_1, y_1), .., (x_{m_{test}},
            y_{m_{test}})\}:$ We use $X_{test}$ to compute and report back the
            final performance results as presented in the evaluation section
            (Section \ref{sec_evaluation}). 

    \item $m_{test} = |X_{test}|:$  $m_{test}$ is the size the $X_{test}$ and
        it represent 50\% of the global dataset $X$.

\end{itemize}

\section{Methodology} \label{sec_approach}

In this section, we present the general approach of MalDy as illustrated in
Figure \ref{fig_dysign_classification_approach_overview}. The section describes
the approach based on the chronological order of the building steps. 

\subsection{Behavioral Reports Generation} \label{sec_report_generation}

MalDy Framework starts from a dataset $X$ of behavioral reports with known
labels. We consider two primary sources for such reports based on the
collection environment. First, We collect the reports from a software sandbox
environment \cite{Toward2017WillemsHF}, in which we execute the binary program,
malware or benign, in a controlled system (mostly virtual machines). The main
usage of sandboxing in security investigation is to check and analyze the
maliciousness of programs. Second, we could collect the behavioral reports from
a production system in the form of system logs of the running apps. The goal is
to investigate the sanity of the apps during the executions and there is no
malicious activity. As presented in Section \ref{sec_overview_intuition}, MalDy
employs a word-based approach to model the behavioral reports, but it is not
clear yet how to use the report's sequence of words in MalDy Framework.


\subsection{Report Vectorization} 
\label{sec_report_vectorization}

In this section, we answer the question: how can we model the words in the
behavioral report to fit in our classification component? Previous solutions
\cite{Sen2016StormDroid} \cite{Automatic2011Rieck} select specific features
from the behavioral reports by: (i) extract relevant security features (ii)
manually inspect and select from these features \cite{Sen2016StormDroid}. This
process involves manual work from the security investigator. Also, it is not
scalable since the investigator needs to redo this process manually for each
new type of behavioral report. In other words, we are looking for features
(words in our case) representation that makes an automatic feature engineering
without the intervention of a security expert. For this purpose, MalDy proposes
to employ Bag of Word (BoW) NLP model. Specifically, we leverage term
frequency-inverse document frequency (TFIDF) \cite{itidf_wiki} or feature
hashing (trick) (FH) \cite{qinfeng09hashk}. MalDy has two variants based on the
chosen BoW technique whether TFIDF or FH. These techniques generate fixed
length vectors for the behavioral reports using words' frequencies. TFIDF and
FH are presented in more detail in Section \ref{sec_framework}. At this point,
we formulate the reports into features vectors, and we are looking to build
classification models. 

\subsection{Build Models} 
\label{sec_build_models}

MalDy framework utilizes a supervised machine learning to build its malware
investigation models. To this point, MalDy is composed of a set of models, each
model has a specific purpose. First, we have the threat detection model that
finds out the maliciousness likelihood of a given app from its behavioral
report. Afterward, the rest machine learning models aim to investigate
individual family threats separately. MalDy uses a model for each possible
threat that the investigator is checking for. In our case, we have a malware
detection model along with a set of malware family attribution models. In this
phase, we build each model separately using $X_{build}$. All the models are
conducting a binary classification to provide the likelihood of a specific
threat. In the process of building MalDy models, we evaluate different
classification algorithms to compare their performance. Furthermore, we tune up
each ML algorithm classification performance under an array of hyper-parameters
(different for each ML algorithm). The latter is a completely automatic
process; the investigator only needs to provide $X_{build}$. We train each
investigation model on $X_{train}$ and tune its performance on $X_{valid}$ by
finding the kptimum algorithm hyper-parameters as presented in Algorithm
\ref{alg_build_models}. Afterward, we determine the optimum decision threshold
for each model using it performance on $X_{valid}$. At the ends this stage, we
have list of optimum models' tuples $Opt = \{<c_0, th_0, params_0>, <c_1, th_1,
params_1>, .. ,<c_c, th_c, params_c>\}$, the cardinality of list $c$ is number
of explored classification algorithms. A tuple $<c_i, th_i, params_i>$ defines
the optimum hyper-parameters $params_i$ and decision threshold $th_i$ for ML
classification algorithm $c_i$.  

\begin{algorithm}
    \SetKwInOut{Input}{Input}
    \SetKwInOut{Output}{Output}

    \Input{$X_{build}$: build set} 
    \Output{$Opt$: optimum models' tuples}
    
    $X_{train}, X_{valid} = X_{build}$
    
    \For{c in MLAlgorithms}{
      score = 0
      \For{params in c.params\_array}{
        model = train($alg, X_{train}, params$) \;
 	s, th = validate(model, $X_{valid}$) \;

	\If{s $>$ score}
	{
	  ct = $<c, th, params>$ \;
	}	
      }
    
      $Opt$.add(ct)
    }
 
    \textbf{return} $Opt$ 

	\caption{Build Models Algorithm}
	\label{alg_build_models}
\end{algorithm}

\subsection{Ensemble Composition} 
\label{sec_ensemble_composition}

Previously, we discuss the process of building and tunning individual
classification model for specific investigation tasks (malware detection,
family one threat attribution, family two threat attribution, etc.). In this
phase, we construct an ensemble model (outperforms single models) from a set of
models generated using the optimum parameters computed previously (Section
\ref{sec_build_models}). We take each set of optimally trained models $\{(C_1,
th_1), (C_2, th_2), .., (C_h, th_h)\}$ for a specific threat investigation task
and unify them into an ensemble $E$. The latter utilizes the majority voting
mechanism between the individual model's outcomes for a specific investigation
task. Equation \ref{equ_ensemble_decision} shows the computation of the final
outcome for one ensemble $E$, where $w_i$ is the weight given for a single
model. The current implementation gives equal weights for the ensemble's
models. We consider exploring $w$ variations for future work. This phase
produces MalDy ensembles, $\{E^{1}_{Detection}, $ $E^{2}_{Family1}, $
$E^{3}_{Family2} $ $.., E^{T}_{familyJ}\}$, an ensemble for each threat and the
outcome is the likelihood this threat to be positive. 


\begin{equation}
\begin{small}
\begin{aligned}
\hat{y} = E(x) &= sign\left(\sum^{|E|}_{i} w_i C_{i}(x, th_{i}) \right) \\
               &= \left\{ 
		   \begin{array}{c} 
                    +1: \sum_{i} \left(w_i C_{i}) \right >= 0 \\
   		    -1: \sum_{i} \left(w_i C_{i}) \right < 0
                   \end{array} 
                  \right.
\end{aligned}
\end{small}
\label{equ_ensemble_decision}
\end{equation}

%

\subsubsection{Ensemble Prediction Process} 
\label{sec_ensemble_prediction}

MalDy prediction process is divided into two phases as depicted in Algorithm
\ref{alg_prediction_algorithm}. First, given a behavioral report, we generate
the feature vector $x$ using TFIDF or FH vectorization techniques. Afterward,
the detection ensemble $E_{detection}$ checks the maliciousness likelihood of
the feature vector $x$. If the maliciousness detection is positive, we proceed
to the family threat attribution. Since the family threat ensembles,
$\{E^{2}_{Family1}, $ $E^{3}_{Family2} $ $.., E^{T}_{familyJ}\}$, are
independent, we compute the outcomes of each family ensemble $E_{family_i}$.
MalDy flags a malware family threat if and only if the majority voting is above
a given voting threshold $vth$ (computed using $X_{valid}$). In the case there is
no family threat flagged by the family ensembles, MalDy will tag the current
sample as an unknown threat. Also, in the case of multiple families are
flagged, MalDy will select the family with the highest probability, and provide
the security investigator with the flagged families sorted by the likelihood
probability. The separation between the family attribution models makes MalDy
more flexible to update. Adding a new family threat will need only to train,
tune, and calibrate the family model without affecting the rest of the
framework ensembles.

%

\begin{algorithm}
    \SetKwInOut{Input}{Input}
    \SetKwInOut{Output}{Output}

    \Input{$report$: Report} 
    \Output{$D$: Decision}

    $E_{detection}$ = $E^{1}_{Detection}$ \;
    $E_{family}$ = $\{E^{2}_{Family1}, .., E^{T}_{familyJ}\}$ \;

    $x$ = Vectorize($report$) \;

    detection\_result = $E_{detection}(x)$\;

    \If{detection\_result $<$ 0}
    {
     \textbf{return} detection\_result \;
    }	

    \For{$E_{F_i}$ in $E_{family}$}{
        family\_result = $E_{F_i}$(x) \; 
    }

    \textbf{return} detection\_result, family\_result \;

\caption{Prediction Algorithm}
\label{alg_prediction_algorithm}
\end{algorithm}


\section{Framework} 
\label{sec_framework}

In this section, we present in more detail the key techniques used in MalDy
framework namely, n-grams \cite{ngram2004AbouAssaleh}, feature hashing (FH),
and term frequency inverse document frequency (TFIDF). Furthermore, we present
the explored and tuned machine learning algorithms during the models building
phase (Section \ref{sec_ml_algs}).

\subsection{Feature Engineering} 
\label{sec_feature_engineering}

In this section, we describe the components of the MalDy related to the
automatic security feature engineering process. 

\subsubsection{Common N-Gram Analysis (CNG)} 
\label{sec_ngrams}

A key tool in MalDy feature engineering process is the common N-gram analysis
(CNG) \cite{ngram2004AbouAssaleh} or simply N-gram. N-gram tool has been
extensively used in text analyses and natural language processing in general
and its applications such as automatic text classification and authorship
attribution \cite{ngram2004AbouAssaleh}. Simply, n-gram computes the contiguous
sequences of n items from a large sequence. In the context of MalDy, we compute
word n-grams on behavioral reports by counting the word sequences of size n.
Notice that the n-grams are extracted using a moving forward window (of size n)
by one step and incrementing the counter of the found feature (word sequence in
the window) by one. The window size n is a hyper-parameter in MalDy framework.
N-gram computation happens simultaneously with the vectorization using FH or
TFIDF  in the form of a pipeline to prevent computation and memory issues due
to the high dimensionality of the n-grams. From a security investigation
perspective, n-grams tool can produce distinguishable features between the
different variations of an event log compared to single word (1-grams)
features. The performance of the malware investigation is highly affected by
the features generated using n-grams (where $n > 0$). Based on BoW model, MalDy
considers the count of unique n-grams as features that will be leveraged by
through a pipeline to the FH or TFIDF.

\subsubsection{Feature Hashing} 
\label{sec_feature_hashing}

The first approach to vectorize the behavioral reports is to employ feature
hashing (FH) \cite{qinfeng09hashk} (also called hashing trick) along with
n-grams. Feature hashing is a machine learning preprocessing technique for
compacting an arbitrary number of features into a fixed-length feature vector.
The feature hashing algorithm, described in Algorithm
\ref{alg_feature_hashing}, takes as input the report N-grams generator and the
target length L of the feature vector. The output is a feature vector $x_i$
with a fixed size L. We normalize $x_i$ using the euclidean norm (also called
L2 norm). As shown in Formula \ref{equ_l2norm}, the euclidean norm is the
square root of the sum of the squared vector values.

\begin{equation} \label{equ_l2norm}
L2Norm(x) =  \|x\|_2 = \sqrt{x^{2}_{1} +.. +  x^{2}_{n}}
\end{equation}

\begin{algorithm}[h]
    \SetKwInOut{Input}{Input}
    \SetKwInOut{Output}{Output}

    \Input{\textbf{X\_seq}: Report Word Sequence, \\
           \textbf{L}: Feature Vector Length}
    \Output{FH: Feature Hashing Vector}
    
    \textbf{ngrams} = Ngram\_Generator(\textbf{X\_seq})\;

    FH = \textbf{new} feature\_vector[\textbf{L}]\;

    \For{\textbf{Item} in \textbf{ngrams}}{
 	H = hash(\textbf{Item}) \;
 	feature\_index = H mod \textbf{L} \;
 	FH[feature\_index] += 1 \;
    }
    // normalization \\
    FH = FH / $\|FH\|_2$ \; 

\caption{Feature Vector Computation}
\label{alg_feature_hashing}
\end{algorithm}

Previous researches \cite{Weinbergeretal09,qinfeng09hashk} have shown that the
hash kernel approximately preserves the vector distance. Also, the
computational cost incurred by using the hashing technique for reducing a
dimensionality grows logarithmically with the number of samples and groups.
Furthermore, it helps to control the length of the compressed vector in an
associated feature space. Algorithm \ref{alg_feature_hashing} illustrates the
overall process of computing the compacted feature vector. 


\subsubsection{Term Frequency-Inverse Document Frequency} 
\label{sec_tfidf}

TFIDF \cite{itidf_wiki} is the second possible approach for behavioral reports
vectorization that also leverages N-grams tool. It is a well-known technique
adopted in the fields of \emph{information retrieval} (IR) and \emph{natural
language processing} (NLP). It computes feature vectors of input behavioral
reports by considering the relative frequency of the n-grams in the individual
reports compared to the whole reports dataset. Let $D=\{d_1, d_2,\ldots, d_n\}$
be a set of behavioral documents, where $n$ is the number of reports, and let
$d=\{w_1, w_2,\ldots, w_m\}$ be a report, where $m$ is the number of n-grams in
$d$. TFIDF  of n-gram $w$ and report $d$ is the product of \emph{term
frequency} of $w$ in $d$ and the \emph{inverse document frequency} of $w$, as
shown in Formula \ref{equ_tfidf}. The \emph{term frequency} (Formula
\ref{equ_tf}) is the occurrence number of $w$ in $d$. Finally, the
\emph{inverse document frequency} of $w$ (Formula \ref{equ_idf}) represents the
number of documents $n$ divided by the number of documents that contain $w$ in
the logarithmic form. Similarly to the feature hashing (Section
\ref{sec_feature_hashing}), we normalize the produced vector using L2 norm (see
    Formula \ref{equ_l2norm}. The computation of TFIDF is very scalable, which
    enhance MalDy efficiency.

\begin{equation}
\mbox{\it tf-idf}(w,d) = \mbox{\it tf}(w,d) \times \mbox{\it idf}(w)
\label{equ_tfidf}
\end{equation}

\begin{equation}
\mbox{\it tf}(w,d) = |w_i \in d, ~d=\{w_1, w_2, ... w_n\}: w = w_i|
\label{equ_tf}
\end{equation}

\begin{equation}
\mbox{\it idf}(w) = log{\frac{|D|}{1 + |d: w \in d|}}
\label{equ_idf}
\end{equation}


\subsection{Machine Learning Algorithms} 
\label{sec_ml_algs}

Table \ref{tab_classifiers_list} shows the candidate machine learning
classification algorithms for MalDy framework. The candidates represent the
most used classification algorithms and come from different learning categories
such as tree-based. Also, all these algorithms have efficient public
implementations. We chose to exclude the logistic regression from the candidate
list due to the superiority of SVM in almost all cases. KNN may consume a lot
of memory resources during the production because it needs all the training
dataset to be deployed in the production environment. However, we keep KNN in
MalDy candidate list because of its unique fast update feature. Updating KNN in
a production environment requires only update the train set, and we do not need
to retrain the model. This option could be very helpful in certain malware
investigation cases. Considering other ML classifiers is considered for future
work design and implementation.

\begin{table}[h]
    \centering
    \begin{scriptsize}
    \begin{tabular}{ccc}
        \hline
        \hline
        Classifier Category & Classifier Algorithm & Chosen \\ 
        \hline

        		& CART 				& \cmark \\ 
        Tree 		& Random Forest 		& \cmark \\ 
        		& Extremely Randomized Trees 	& \cmark \\
        \hline
        General  	& K-Nearest Neighbor (KNN) 	& \cmark \\  
          		& Support Vector Machine (SVM)  & \cmark \\ 
          		& Logistic Regression 		& \xmark \\ 
        	 	& XGBoost 			& \cmark \\ 
        \hline
        \hline
    \end{tabular}
    \end{scriptsize}
    \caption{Explored Machine Learning Classifiers}
    \label{tab_classifiers_list}
\end{table}


%


%

\section{Evaluation Datasets} 
\label{sec_dataset}

Table \ref{tab_dataset_list} presents the different datasets used to evaluate
MalDy framework. We focus on the Android and Win32 platforms to prove the
portability of MalDy, and other platforms are considered for a further future
research. All the used datasets are publicly available except the Win32 Malware
dataset, which is  provided by a third-party security vendor. The behavioral
reports are generated using DroidBox \cite{droidbox_github} and ThreatAnalayzer
\footnote{threattrack.com} for Android and Win32 respectively.

\begin{table}[h]
\centering
\begin{scriptsize}
\begin{tabular}{ccccc}
\hline
\hline
Platform& Dataset		& Sandbox 	& Tag 	& \#Sample/\#Family \\ 
\hline                                                                
	& MalGenome \cite{zhou2012dissecting} & D   & Malware &  1k/10       \\ 
Android & Drebin    \cite{arp2014drebin}      & D   & Malware &  5k/10       \\ 
	& Maldozer  \cite{karbab2018maldozer} & D   & Malware & 20k/20       \\ 
	& AndroZoo  \cite{Allix2016AndroZoo}  & D   & Benign  & 15k/-        \\ 
	& PlayDrone \footnote{https://archive.org/details/android\_apps} 
           & D   & Benign  & 15k/-        \\ 
\hline                                                   
Win32 	& Malware   \footnote{https://threattrack.com/} & T   
           & Malware & 20k/15       \\ 
\hline
\hline
\end{tabular}
\end{scriptsize}
\caption{Evaluation Datasets. D: DroidBox, T: ThreatAnalyzer}
\label{tab_dataset_list}
\end{table}

\section{MalDy Evaluation} 
\label{sec_evaluation}

\begin{figure*}[h]
\begin{center}
\subfigure[\scriptsize General]{%
    \label{fig_maldy_overall_performance}
    \includegraphics[width=.35\textwidth]{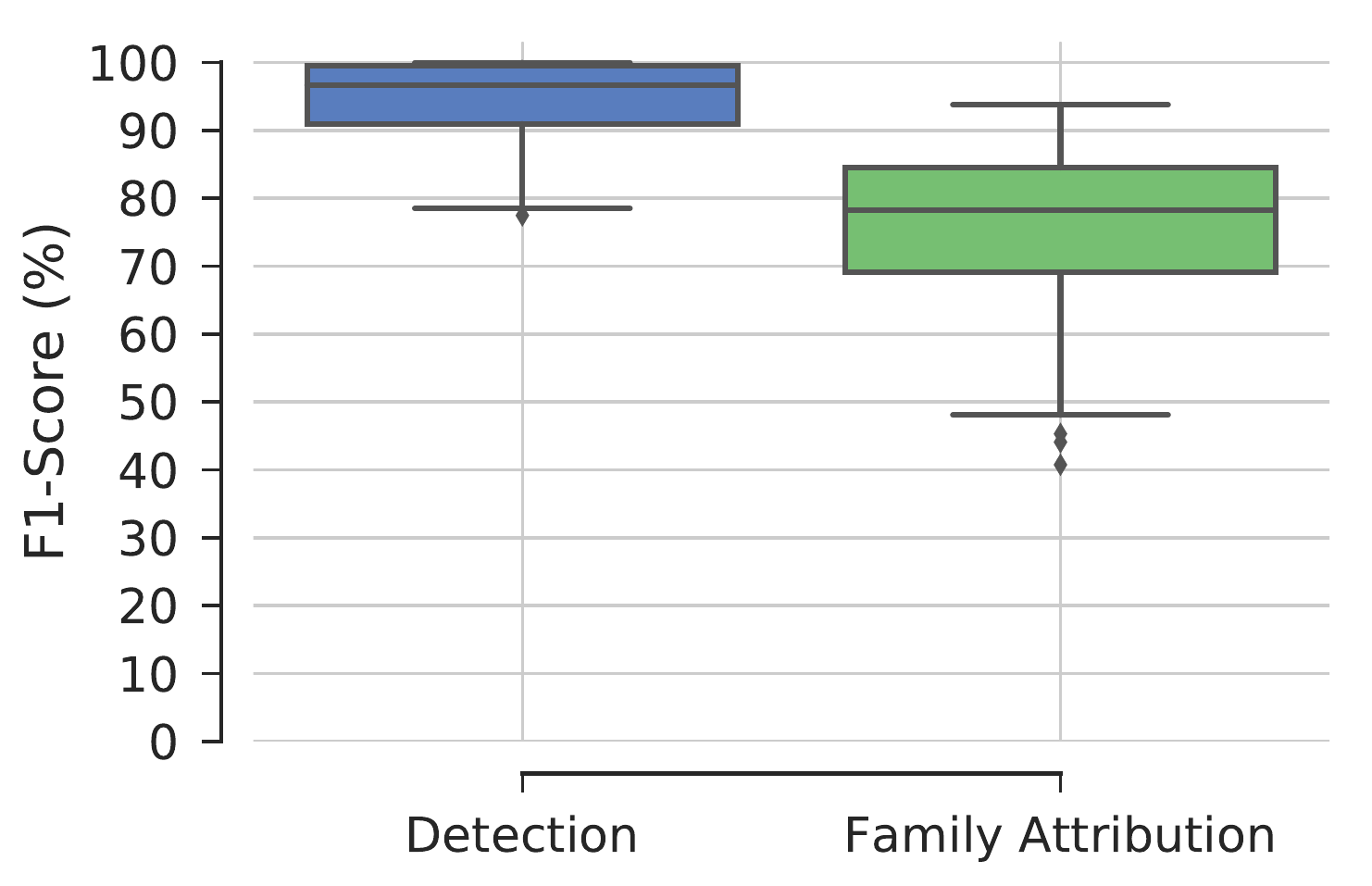}
}%
\subfigure[\scriptsize Malgenome]{%
    \label{fig_maldy_genome_performance}
    \includegraphics[width=.35\textwidth]{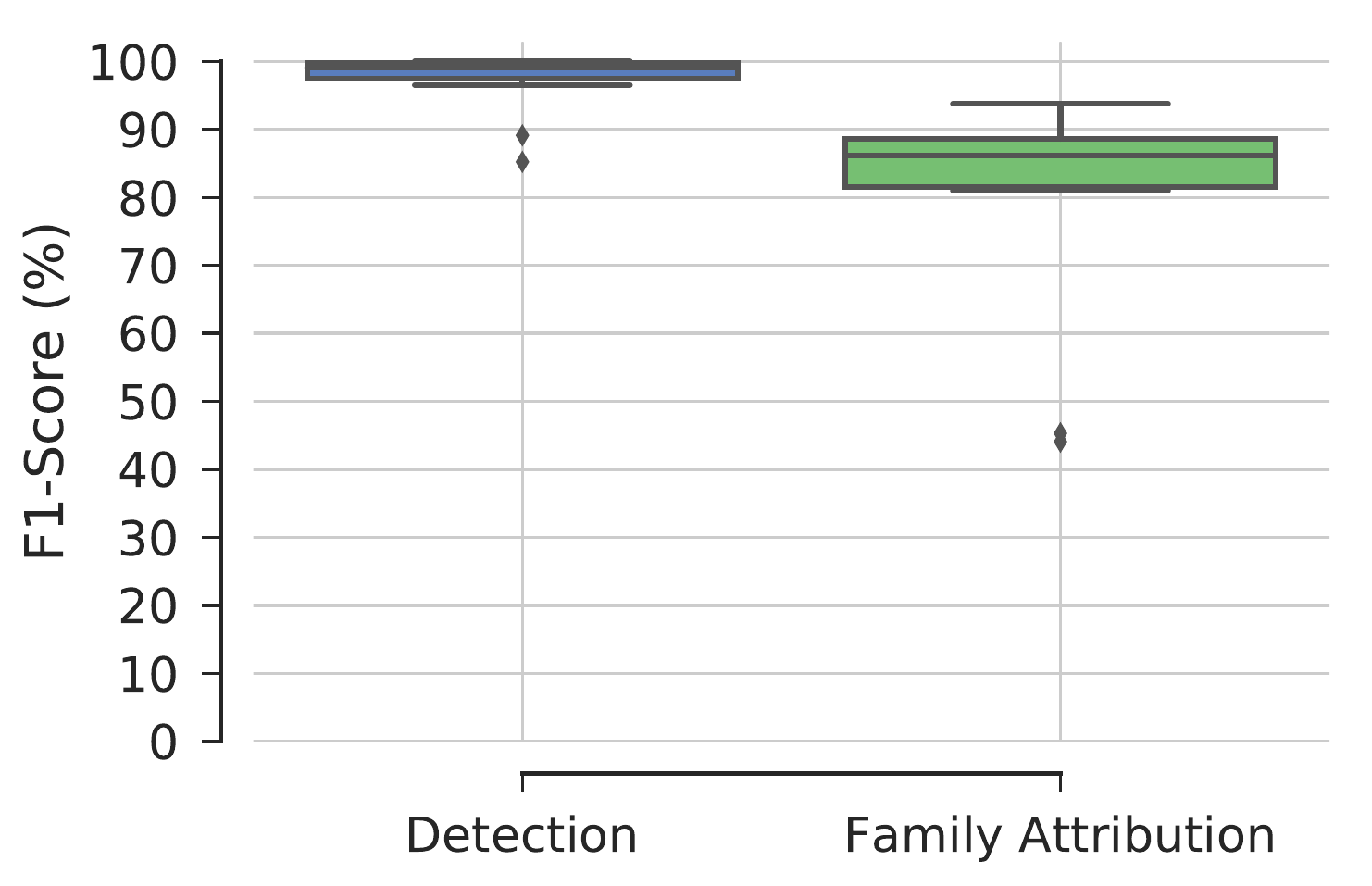}
}\\%
\subfigure[\scriptsize Drebin]{%
    \label{fig_maldy_drebin_performance}
    \includegraphics[width=.35\textwidth]{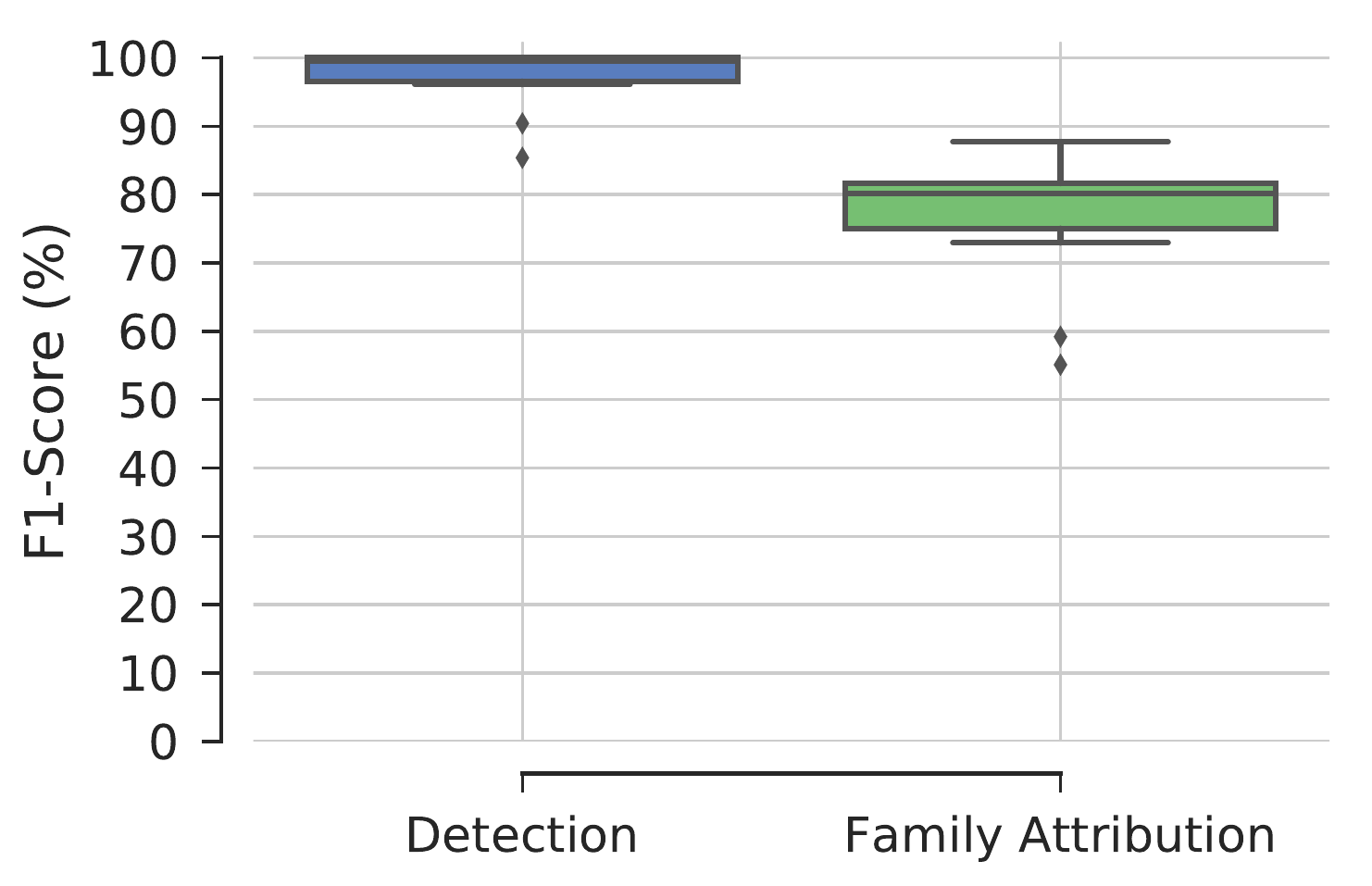}
}%
\subfigure[\scriptsize Maldozer]{%
    \label{fig_maldy_maldozer_performance}
    \includegraphics[width=.35\textwidth]{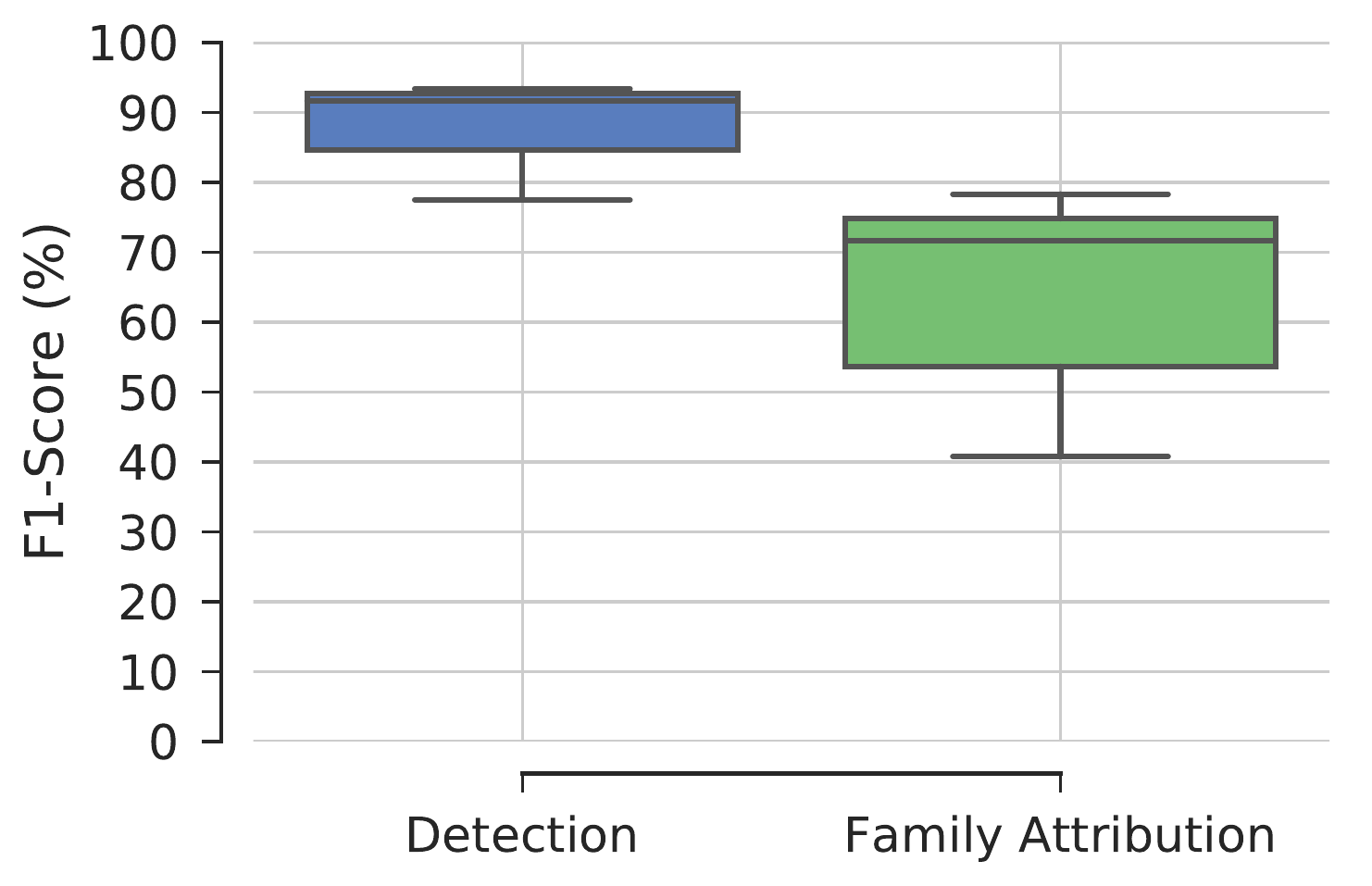}
}%
\end{center}
\caption{MalDy Effectiveness Performance}
\label{fig_maldy_effectiveness}
\end{figure*}

In the section, we evaluate MalDy framework on different datasets and various
settings. Specifically, we question the effectiveness of the word-based
approach for malware detection and family attribution on Android behavior
reports (Section \ref{sec_evaluation_effectiveness}). We verify the portability
and MalDy concept on other platforms (Win32 malware) behavioral reports
(Section \ref{sec_evaluation_portablity}). Finally, We measure the efficiency
of MalDy under different machine learning classifiers and vectorization
techniques (Section \ref{sec_evaluation_efficiency}). During the evaluation, we
answer some other questions related to the comparison between the vectorization
techniques (Section \ref{sec_eval_vectorization_effect}, and the used
classifiers in terms of effectiveness and efficiency (Section
\ref{sec_eval_classifier_effect}). Also, we show the effect of train-set's
size (Section \ref{sec_eval_maldy_train_size}) and the usage of machine
learning ensemble technique (Section \ref{sec_eval_tuning_effect}) on the
framework performance.

\subsection{Effectiveness} 
\label{sec_evaluation_effectiveness}

The most important question in this research is: Can MalDy framework detect
malware and make family attribution using a word-based model on behavioral
reports? In other words, how effective this approach? Figure
\ref{fig_maldy_effectiveness} shows the detection and the attribution
performance under various settings and datasets. The settings are the used
classifiers in ML ensembles and their hyper-parameters, as shown in Table
\ref{tab_Android_detection}. Figure \ref{fig_maldy_overall_performance} depicts
the overall performance of MalDy. In the detection, MalDy achieves 90\%
f1-score (100\% maximum and about 80\% minimum) in most cases. On the other
hand, in the attribution task, MalDy shows over 80\% f1-score in the various
settings. More granular results for each dataset are shown in Figures
\ref{fig_maldy_genome_performance}, \ref{fig_maldy_drebin_performance}, and
\ref{fig_maldy_maldozer_performance} for Malgenome \cite{zhou2012dissecting},
Drebin \cite{arp2014drebin}, and Maldozer \cite{karbab2018maldozer} datasets
respectively. Notice that Figure \ref{fig_maldy_overall_performance} combines
the performance of based (worst), tuned, and ensemble models, and summaries the
results in Table \ref{tab_tuning_effect}.

\begin{table}[h!]
\centering 
\begin{tabular}{l|ccc|ccc}
\hline
\hline
 	  & \multicolumn{3}{c}{\textbf{Detection (F1 \%)}}&\multicolumn{3}{c}{\textbf{Attribution (F1 \%)}}  \\
 	  & \textbf{Base}   & \textbf{Tuned}  & \textbf{Ens}    & \textbf{Base}   & \textbf{Tuned}  & \textbf{Ens}         \\ 
\hline
\hline
\textbf{General}   &        &        &        &        &        &    \\
mean 	  & 86.06  & 90.47  & 94.21  & 63.42  & 67.91  & 73.82       \\
std 	  &  6.67  &  6.71  &  6.53  & 15.94  & 15.92  & 14.68       \\
min 	  & 69.56  & 73.63  & 77.48  & 30.14  & 34.76  & 40.75       \\
25\%	  & 83.58  & 88.14  & 90.97  & 50.90  & 55.58  & 69.07       \\
50\%	  & 85.29  & 89.62  & 96.63  & 68.81  & 73.31  & 78.21       \\
75\%	  & 91.94  & 96.50  & 99.58  & 73.60  & 78.07  & 84.52       \\
max 	  & 92.81  & 97.63  & 100.0  & 86.09  & 90.41  & 93.78       \\
\hline
\textbf{Genome}    &        &        &        &        &        &    \\
mean      & 88.78  & 93.23  & 97.06  & 71.19  & 75.67  & 79.92       \\
std       &  5.26  &  5.46  &  4.80  & 16.66  & 16.76  & 16.81       \\
min       & 77.46  & 81.69  & 85.23  & 36.10  & 40.10  & 44.09       \\
25\%      & 85.21  & 89.48  & 97.43  & 72.36  & 77.03  & 81.47       \\
50\%      & 91.82  & 96.29  & 99.04  & 76.66  & 81.46  & 86.16       \\
75\%      & 92.13  & 96.68  & 99.71  & 80.72  & 84.82  & 88.61       \\
max       & 92.81  & 97.63  & 100.0  & 86.09  & 90.41  & 93.78       \\
\hline
\textbf{Drebin}    &        &        &        &        &        &    \\
mean      & 88.92  & 93.34  & 97.18  & 65.97  & 70.37  & 76.47       \\
std       &  4.93  &  4.83  &  4.65  &  9.23  &  9.14  &  9.82       \\
min       & 78.36  & 83.35  & 85.37  & 47.75  & 52.40  & 55.10       \\
25\%      & 84.95  & 89.34  & 96.56  & 61.67  & 65.88  & 75.05       \\
50\%      & 91.60  & 95.86  & 99.47  & 69.62  & 74.30  & 80.16       \\
75\%      & 92.25  & 96.53  & 100.0  & 72.68  & 76.91  & 81.61       \\
max       & 92.78  & 97.55  & 100.0  & 76.28  & 80.54  & 87.71       \\
\hline
\textbf{Maldozer}  &        &        &        &        &        &    \\
mean      & 80.48  & 84.85  & 88.38  & 53.11  & 57.68  & 65.06       \\
std       &  6.22  &  6.20  &  5.95  & 16.03  & 15.99  & 13.22       \\
min       & 69.56  & 73.63  & 77.48  & 30.14  & 34.76  & 40.75       \\
25\%      & 75.69  & 80.13  & 84.56  & 39.27  & 43.43  & 53.65       \\
50\%      & 84.20  & 88.68  & 91.58  & 56.62  & 61.03  & 71.65       \\
75\%      & 84.88  & 89.01  & 92.72  & 67.34  & 71.89  & 74.78       \\
max       & 85.68  & 89.97  & 93.39  & 71.17  & 76.04  & 78.30       \\
\hline
\hline
\end{tabular}
\caption{Tuning Effect of Tuning of MalDy Performance}
\label{tab_tuning_effect}
\end{table}                        	


\subsubsection{Classifier Effect}  
\label{sec_eval_classifier_effect}

\begin{figure}[H]
\begin{center}
\subfigure[\scriptsize General]{%
    \label{fig_maldy_classifier_performance}
    \includegraphics[width=.3\textwidth]{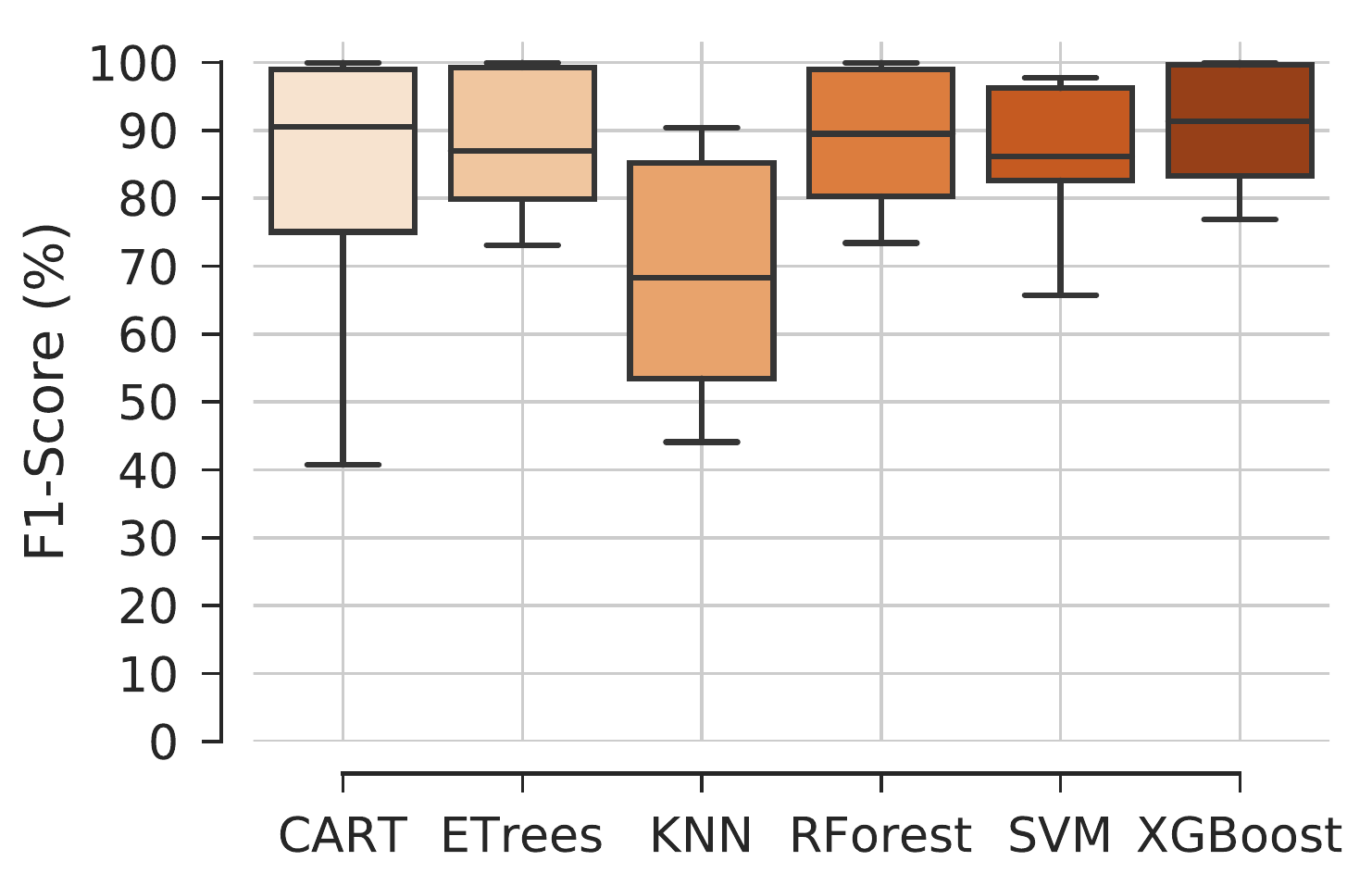}
}\\%
\subfigure[\scriptsize Detection]{%
    \label{fig_detection_classifier_performance}
    \includegraphics[width=.3\textwidth]{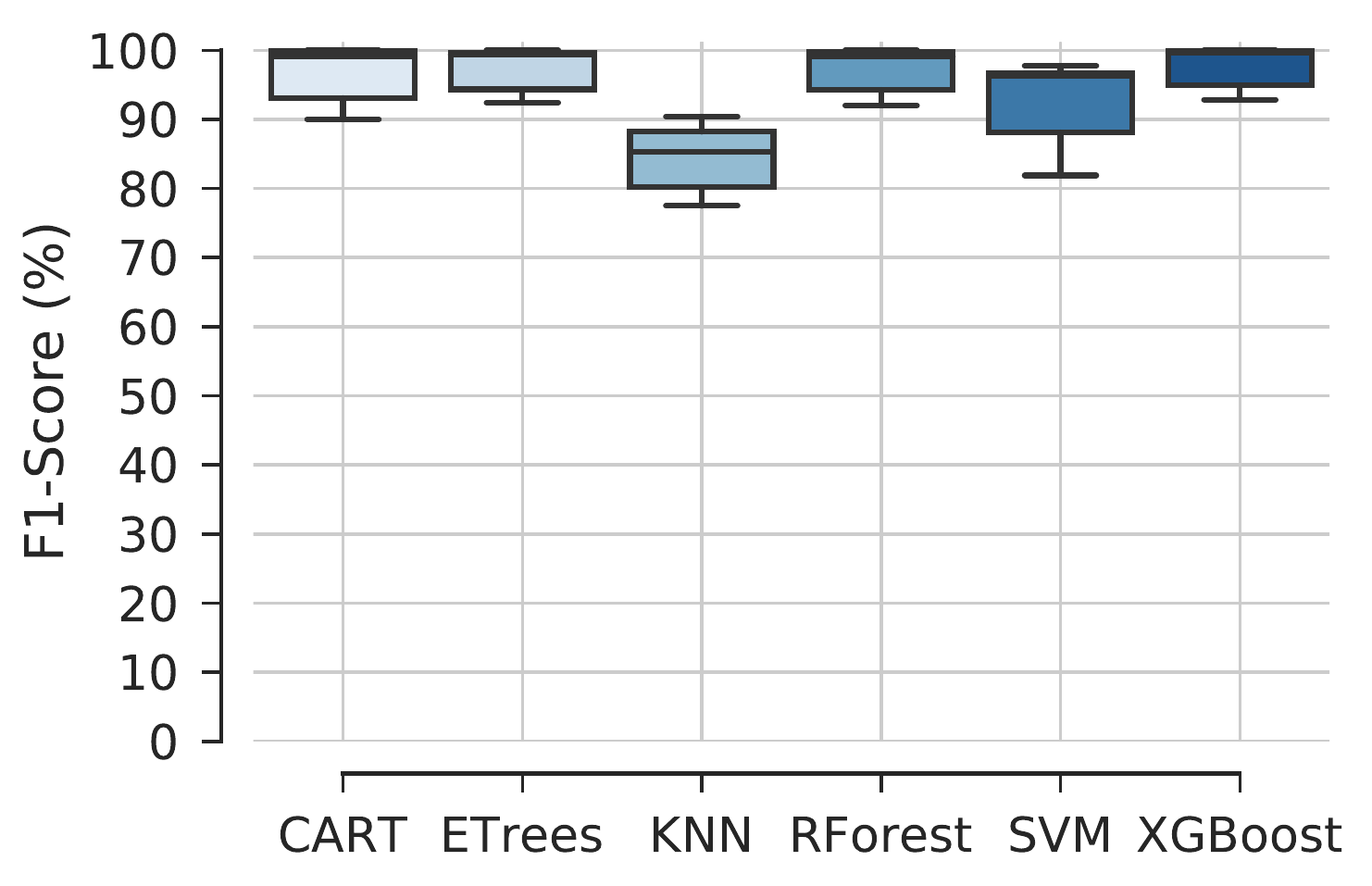}
}\\%
\subfigure[\scriptsize Attribution]{%
    \label{fig_attribution_classifier_performance}
    \includegraphics[width=.3\textwidth]{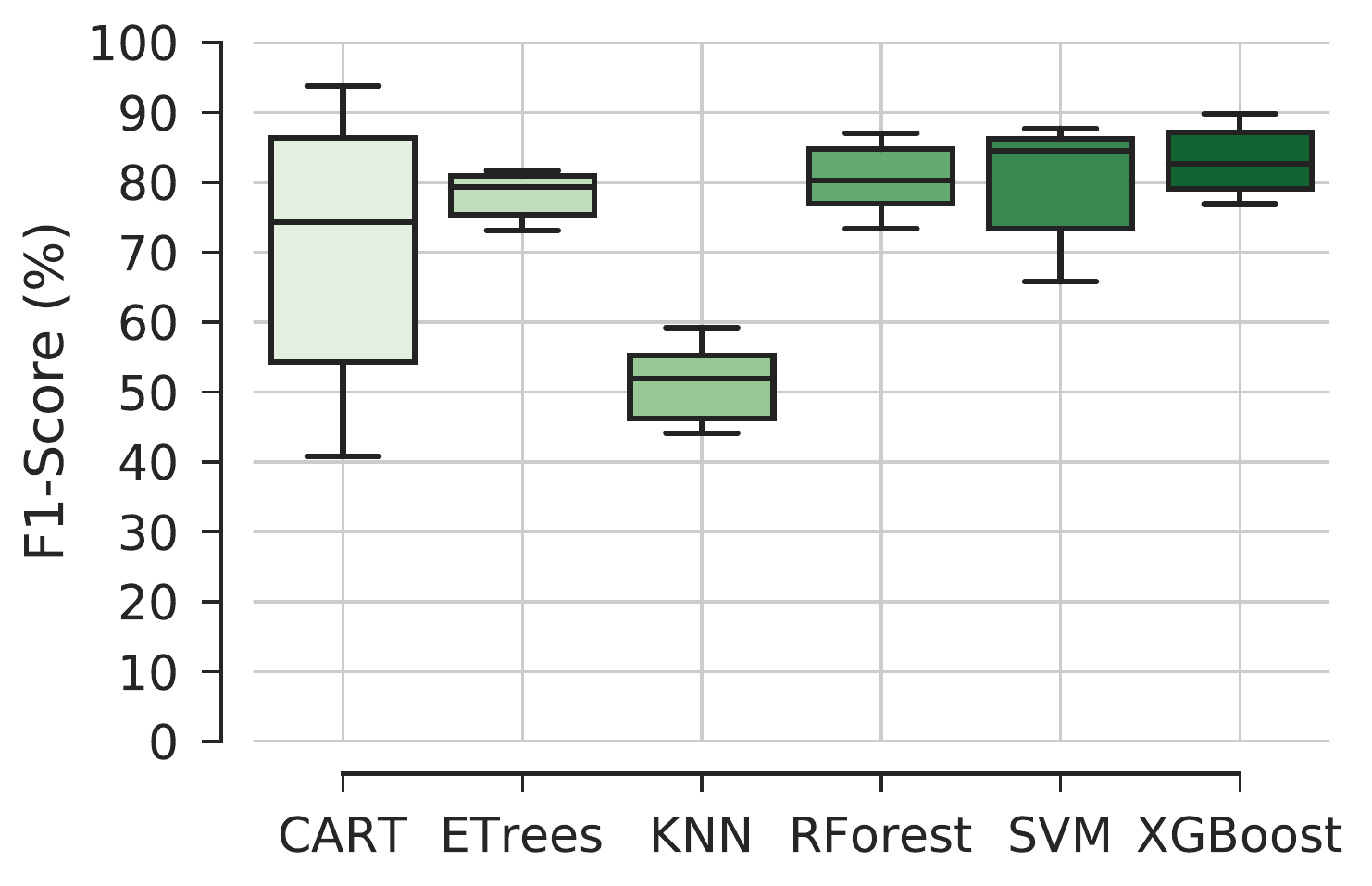}
}%
\end{center}
\caption{MalDy Effectiveness per Machine Learning Classifier}
\label{fig_maldy_effectiveness_ml_classifier}
\end{figure}

The results in Figure \ref{fig_maldy_effectiveness_ml_classifier}, Table
\ref{tab_tuning_effect}, and the detailed Table \ref{tab_Android_detection}
confirm the effectiveness of MalDy framework and its word-based approach.
Figure \ref{fig_maldy_effectiveness_ml_classifier} presents the effectiveness
performance of MalDy using the different classifier for the final ensemble
models. Figure \ref{fig_maldy_classifier_performance} shows the combined
performance of the detection and attribution in f1-score. All the ensembles
achieved a good f1-score, and XGBoost ensemble shows the highest scores. Figure
\ref{fig_detection_classifier_performance} confirms the previous notes for the
detection task. Also, Figure \ref{fig_attribution_classifier_performance}
presents the malware family attribution scores per ML classifier. More details
on the classifiers performance is depicted in Table
\ref{tab_Android_detection}.


\subsubsection{Vectorization Effect} 
\label{sec_eval_vectorization_effect}

\begin{figure}[H]
\begin{center}
\subfigure[\scriptsize General]{%
    \label{fig_maldy_preprocessing_overall_performance}
    \includegraphics[width=.4\textwidth]
        {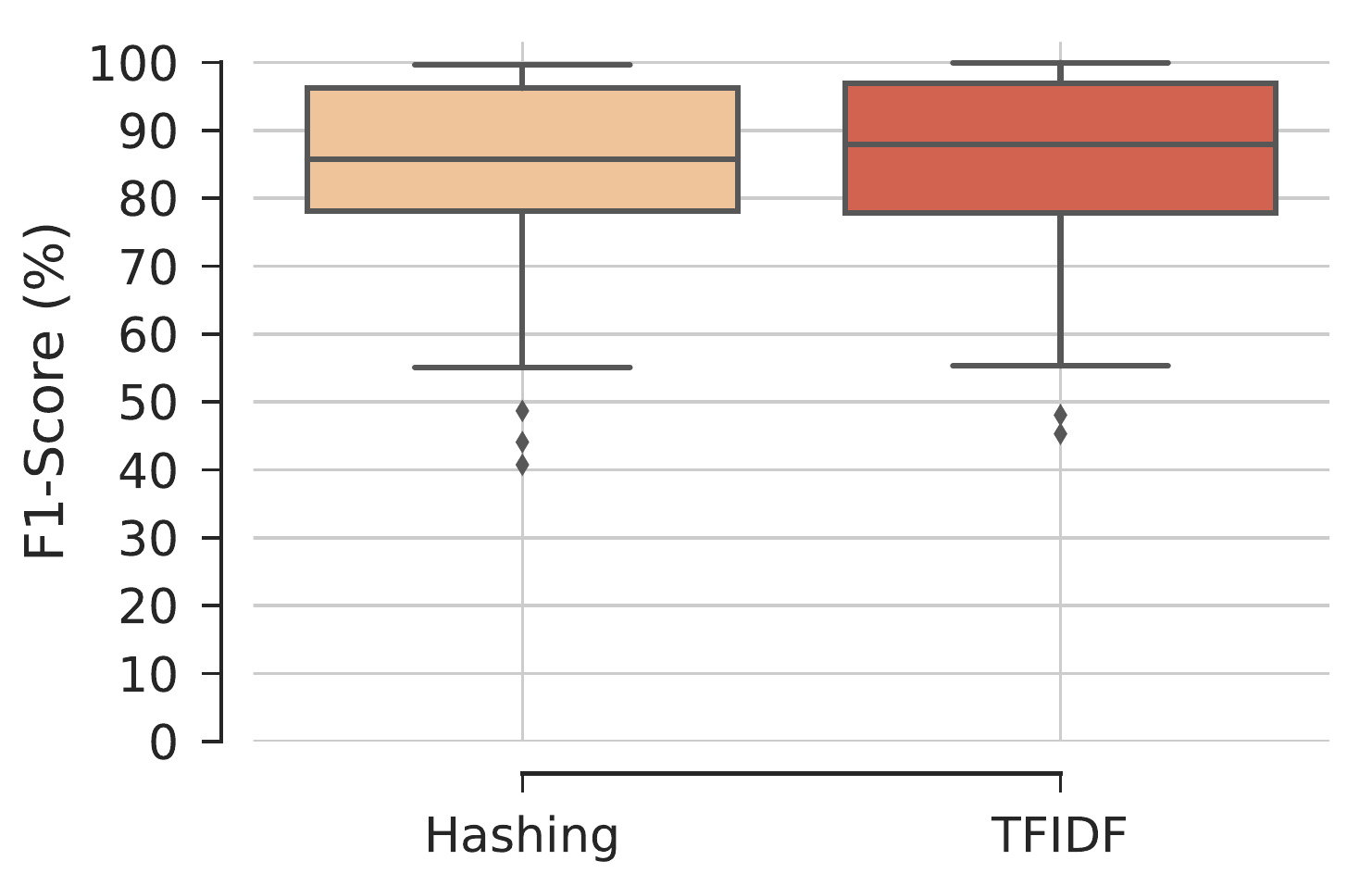}
}\\%
\subfigure[\scriptsize Detection]{%
    \label{fig_detection_preprocessing_performance}
    \includegraphics[width=.4\textwidth]
        {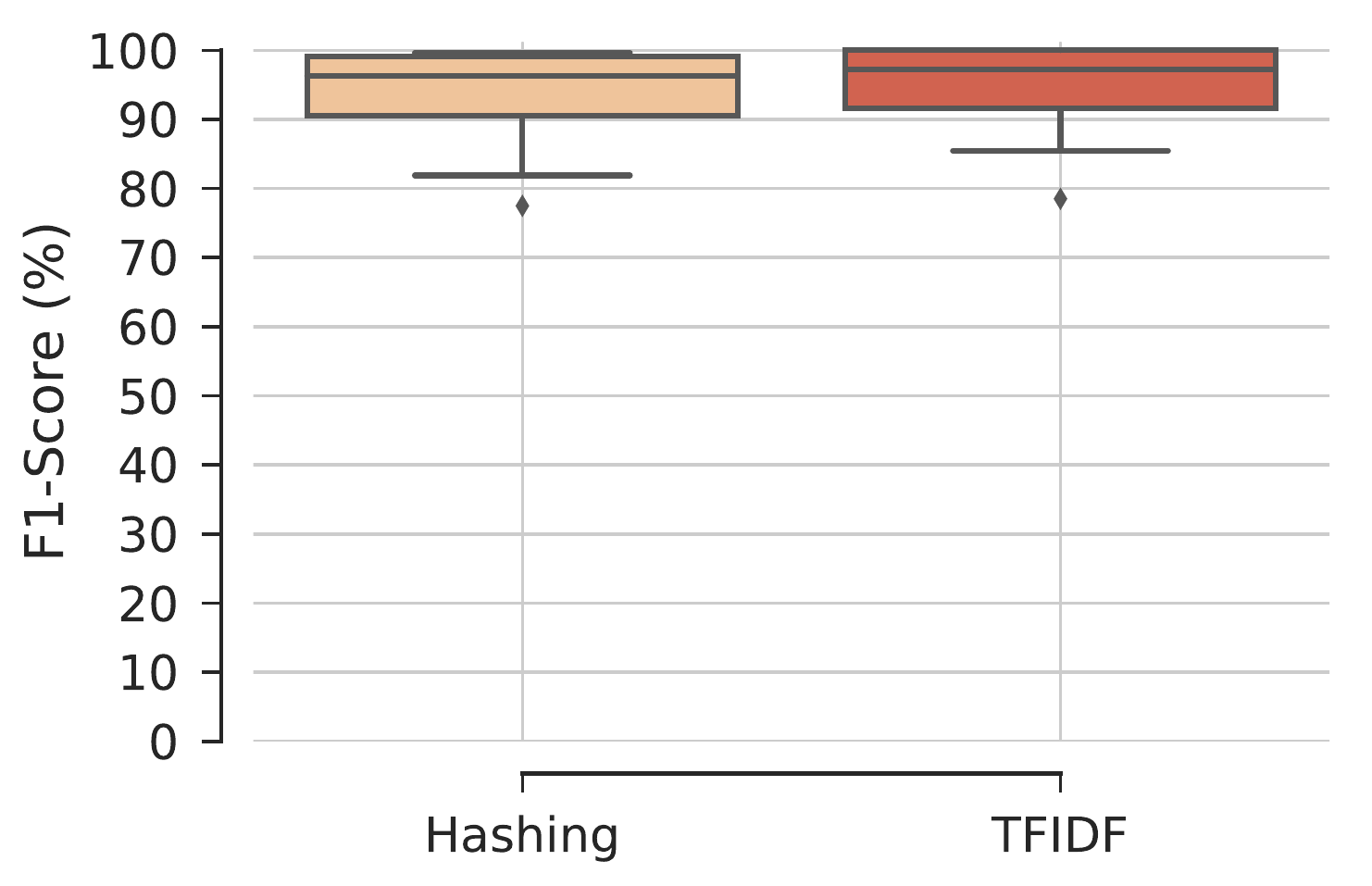}
}\\%
\subfigure[\scriptsize Attribution]{%
    \label{fig_attribution_preprocessing_performance}
    \includegraphics[width=.4\textwidth]
        {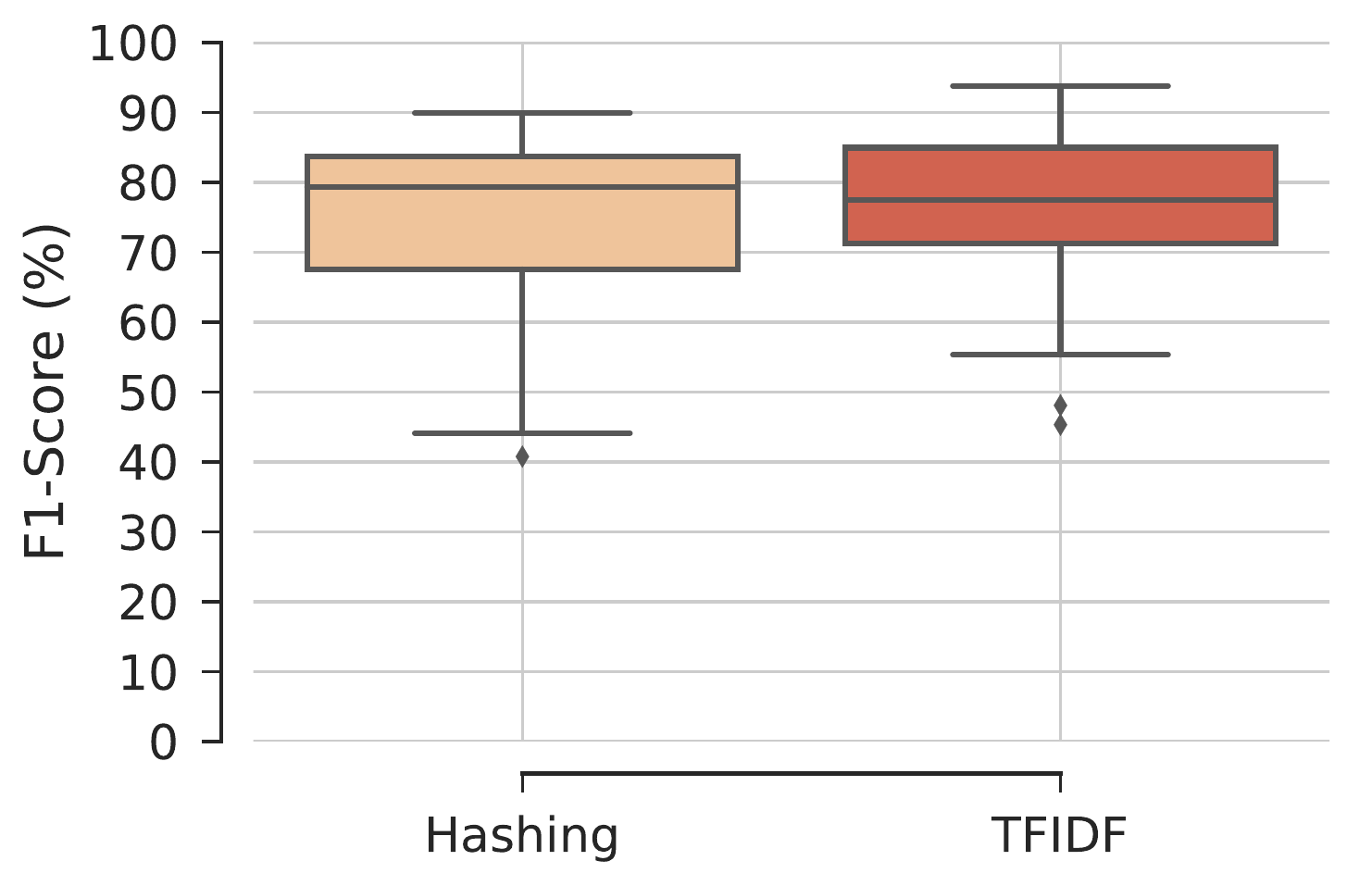}
}%
\end{center}
\caption{MalDy Effectiveness per Vectorization Technique}
\label{fig_maldy_effectiveness_vectorization_technique}
\end{figure}

Figure \ref{fig_maldy_effectiveness_vectorization_technique} shows the effect
of vectorization techniques on the detection and the attribution performance.
Figure \ref{fig_maldy_preprocessing_overall_performance} depict the overall
combined performance under the various settings. As depicted in Figure
\ref{fig_maldy_preprocessing_overall_performance}, Feature hashing and TFIDF
show a very similar performance. In detection task, the vectorization
techniques' f1-score is almost identical as presented in Figure
\ref{fig_detection_preprocessing_performance}. We notice a higher overall
attribution score using TFIDF compared to FH, as shown in Figure
\ref{fig_attribution_preprocessing_performance}. However, we may have cases
where FH outperforms TFIDF. For instance, XGBoost achieved a higher attribution
score under the feature hashing vectorization, as shown in Table
\ref{tab_Android_detection}. 

\begin{table*}[h!]
\centering
\begin{tabular}{lcc|ccc|cccc}
\hline
\hline
\multicolumn{3}{c}{Settings} & \multicolumn{3}{|c|}{\textbf{Attribution} F1-Score (\%)} & \multicolumn{4}{c}{\textbf{Detection} F1-Score (\%) } \\
\hline
\textbf{Model} & \textbf{Dataset} & \textbf{Vector} & \textbf{Base} & \textbf{Tuned} & \textbf{Ensemble} & \textbf{Base} & \textbf{Tuned} & \textbf{Ensemble} & \textbf{FPR}(\%) \\
\hline                          
\hline                          
CART		 & Drebin   & Hashing 	& 64.93 & 68.94 & 	   72.92  & 91.55 & 95.70 & 	    99.40  & 00.64     \\
        	 & Drebin   & TFIDF 	& 68.12 & 72.48 & 	   75.76  & 92.48 & 96.97 & 	    100.0  & 00.00     \\
        	 & Genome   & Hashing 	& 82.59 & 87.28 & 	   89.90  & 91.79 & 96.70 & 	    98.88  & 00.68     \\
        	 & Genome   & TFIDF 	& 86.09 & 90.41 & 	   93.78  & 92.25 & 96.50 & 	    100.0  & 00.00     \\
        	 & Maldozer & Hashing 	& 33.65 & 38.56 & 	   40.75  & 82.59 & 87.18 & 	    90.00  & 06.92     \\
        	 & Maldozer & TFIDF 	& 40.14 & 44.21 & 	   48.07  & 83.92 & 88.67 & 	    91.16  & 04.91     \\
\hline                          
ETrees 		 & Drebin   & Hashing 	& 72.84 & 77.27 & 	   80.41  & 91.65 & 95.77 & 	    99.54  & 00.23     \\
    		 & Drebin   & TFIDF 	& 71.12 & 76.12 & 	   78.13  & 92.78 & 97.55 & 	    100.0  & 00.00     \\
    		 & Genome   & Hashing 	& 74.41 & 79.20 & 	   81.63  & 91.91 & 96.68 & 	    99.14  & 00.16     \\
    		 & Genome   & TFIDF 	& 73.83 & 78.65 & 	   81.02  & 92.09 & 96.61 & 	    99.57  & 00.03     \\
    		 & Maldozer & Hashing 	& 65.23 & 69.34 & 	   73.13  & 84.56 & 88.70 & 	    92.42  & 06.53     \\
    		 & Maldozer & TFIDF 	& 67.14 & 71.85 & 	   74.42  & 84.84 & 88.94 & 	    92.74  & 06.41     \\
\hline                           
KNN 		 & Drebin   & Hashing 	& 47.75 & 52.40 & 	   55.10  & 78.36 & 83.35 & 	    85.37  & 12.86     \\
 		 & Drebin   & TFIDF 	& 51.87 & 56.53 & 	   59.20  & 82.48 & 86.57 & 	    90.40  & 05.83     \\
 		 & Genome   & Hashing 	& 36.10 & 40.10 & 	   44.09  & 77.46 & 81.69 & 	    85.23  & 07.01     \\
 		 & Genome   & TFIDF 	& 37.66 & 42.01 & 	   45.31  & 81.22 & 85.30 & 	    89.13  & 02.10     \\
 		 & Maldozer & Hashing 	& 41.68 & 46.67 & 	   48.69  & 69.56 & 73.63 & 	    77.48  & 26.21     \\
 		 & Maldozer & TFIDF 	& 48.02 & 52.73 & 	   55.31  & 70.94 & 75.36 & 	    78.51  & 03.86     \\
\hline                           
RForest 	 & Drebin   & Hashing 	& 72.63 & 76.80 & 	   80.46  & 91.54 & 95.95 & 	    99.12  & 00.99     \\
		 & Drebin   & TFIDF 	& 72.15 & 76.40 & 	   79.91  & 92.31 & 96.62 & 	    100.0  & 00.00     \\
		 & Genome   & Hashing 	& 78.92 & 83.73 & 	   86.12  & 91.37 & 95.79 & 	    98.95  & 00.68     \\
		 & Genome   & TFIDF 	& 79.45 & 83.90 & 	   87.00  & 92.75 & 97.49 & 	    100.0  & 00.00     \\
		 & Maldozer & Hashing 	& 66.06 & 70.72 & 	   73.41  & 84.49 & 88.96 & 	    92.01  & 07.37     \\
		 & Maldozer & TFIDF 	& 67.96 & 72.04 & 	   75.89  & 85.07 & 89.41 & 	    92.72  & 06.10     \\
\hline                           
SVM 		 & Drebin   & Hashing 	& 57.35 & 61.95 & 	   82.92  & 84.50 & 89.33 & 	    96.08  & 00.86     \\
 		 & Drebin   & TFIDF 	& 63.11 & 67.19 & 	   87.71  & 85.11 & 89.35 & 	    96.73  & 01.15     \\
 		 & Genome   & Hashing 	& 69.99 & 74.68 & 	   86.08  & 85.47 & 89.83 & 	    96.54  & 00.19     \\
 		 & Genome   & TFIDF 	& 73.16 & 77.82 & 	   86.20  & 84.46 & 88.46 & 	    97.73  & 00.39     \\
 		 & Maldozer & Hashing 	& 30.14 & 34.76 & 	   65.76  & 72.32 & 77.12 & 	    81.88  & 15.82     \\
 		 & Maldozer & TFIDF 	& 36.69 & 41.09 & 	   70.18  & 76.82 & 81.14 & 	    85.46  & 08.56     \\
\hline                           
\textbf{XGBoost} & Drebin   & Hashing 	& 76.28 & 80.54 & \textbf{84.01} & 92.05 & 96.50 & \textbf{99.61} & \textbf{00.29}     \\
\textbf{} 	 & Drebin   & TFIDF 	& 73.53 & 77.88 & \textbf{81.18} & 92.23 & 96.45 & \textbf{100.0} & \textbf{00.00}     \\
\textbf{} 	 & Genome   & Hashing 	& 81.80 & 85.84 & \textbf{89.75} & 91.86 & 96.09 & \textbf{99.62} & \textbf{00.32}     \\
\textbf{} 	 & Genome   & TFIDF 	& 80.36 & 84.48 & \textbf{88.24} & 92.81 & 97.63 & \textbf{100.0} & \textbf{00.00}     \\
\textbf{} 	 & Maldozer & Hashing 	& 71.17 & 76.04 & \textbf{78.30} & 85.68 & 89.97 & \textbf{93.39} & \textbf{05.86}     \\
\textbf{} 	 & Maldozer & TFIDF 	& 69.51 & 74.15 & \textbf{76.87} & 85.01 & 89.16 & \textbf{92.86} & \textbf{06.05}     \\
\hline
\hline
\end{tabular}
\caption{Android Malware Detection}
\label{tab_Android_detection}
\end{table*}                        	

\subsubsection{Tuning Effect} 
\label{sec_eval_tuning_effect}

\begin{figure}[h!]
\begin{center}
\subfigure[\scriptsize Detection]{%
    \label{fig_detection_tunning_performance}
    \includegraphics[width=0.35\textwidth]
        {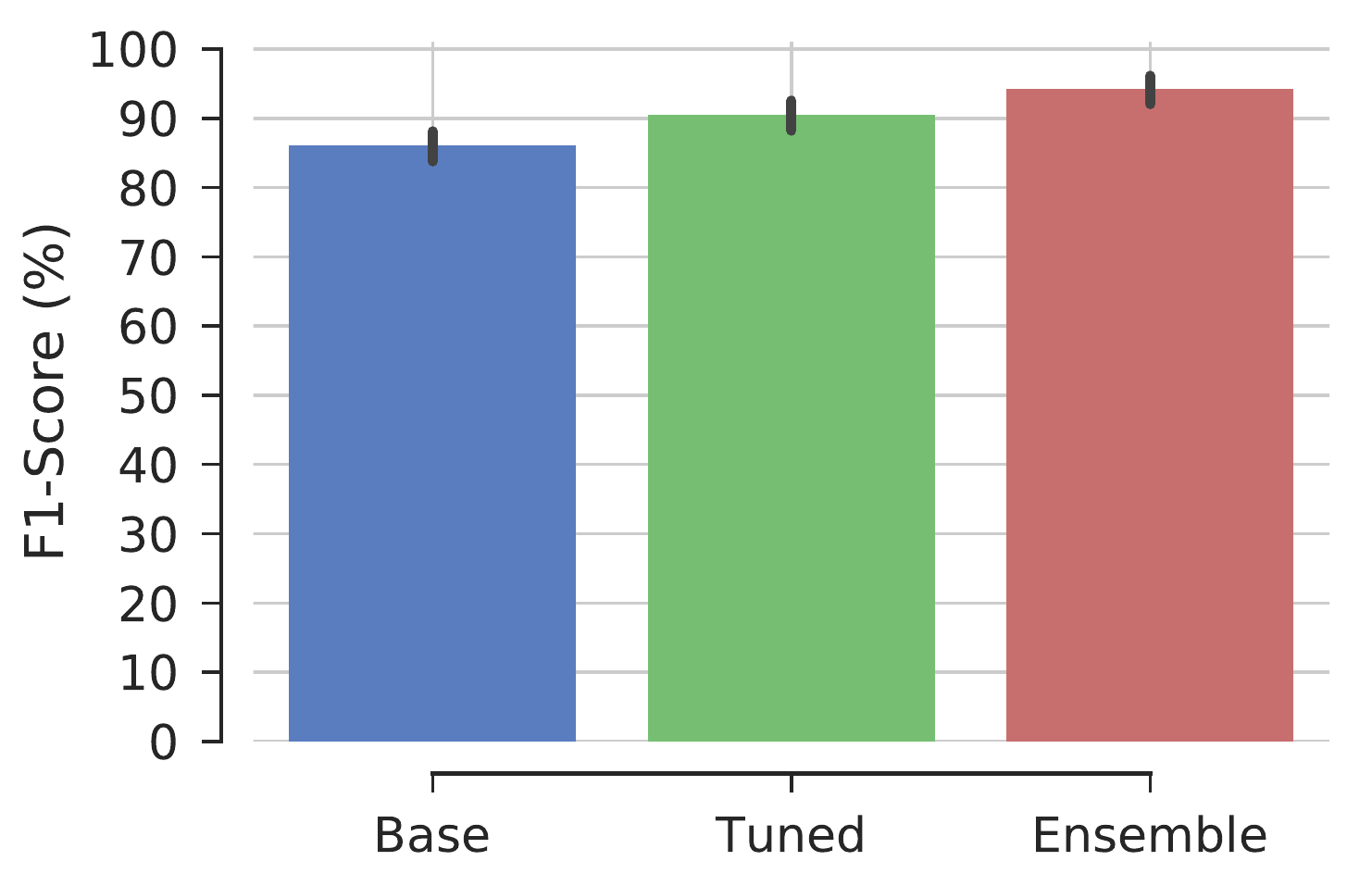}
}\\%
\subfigure[\scriptsize Attribution]{%
    \label{fig_attribution_tunning_performance}
    \includegraphics[width=0.35\textwidth]
        {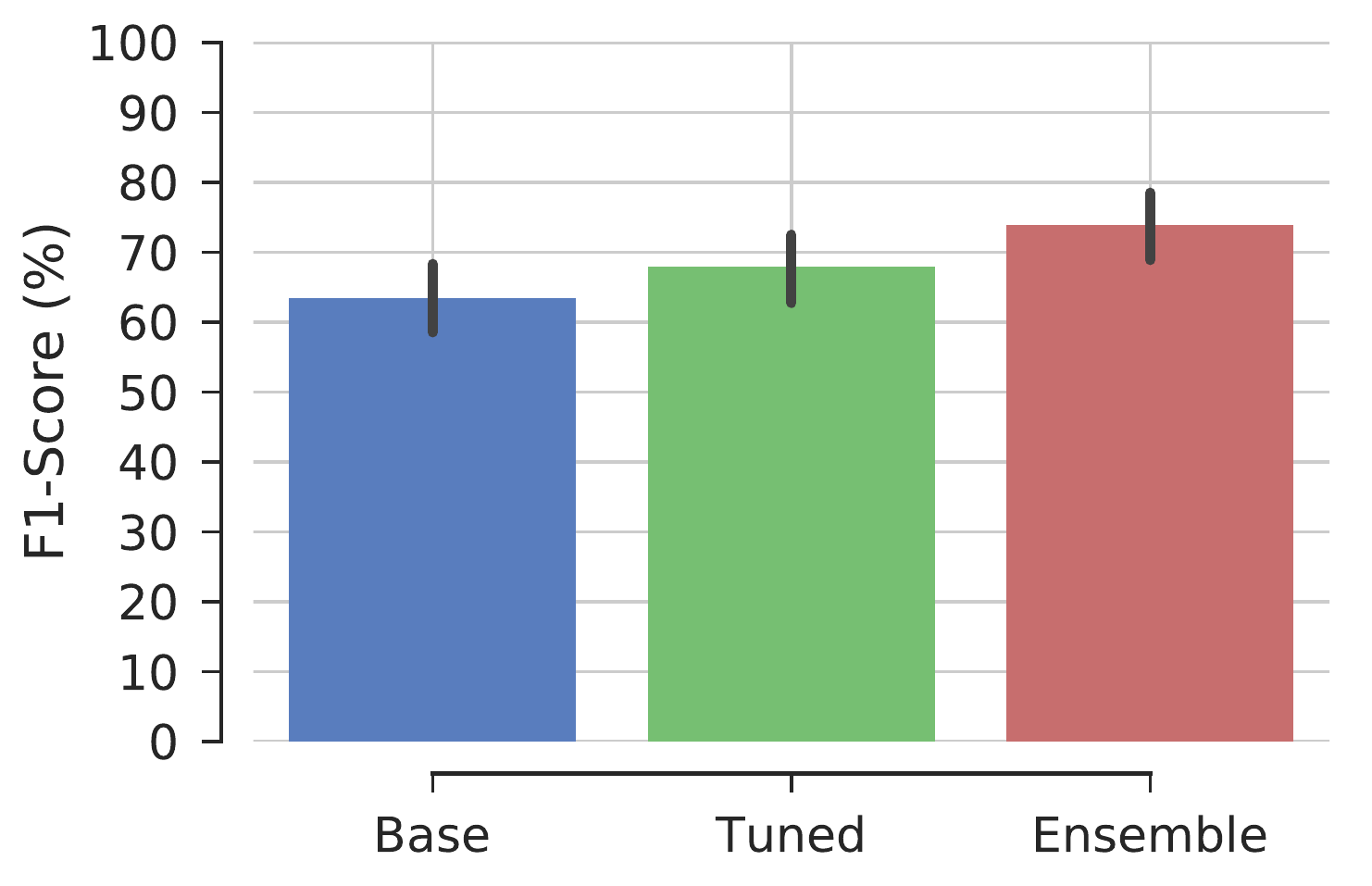}
}%
\end{center}
\caption{Effect of MalDy Ensemble and Tunning on the Performance}
\label{fig_maldy_effectiveness_ens_classifier}
\end{figure}

Figure \ref{fig_maldy_effectiveness_ens_classifier} illustrates the effect of
tune and ensemble phases on the overall performance of MalDy. In the detection
task, as in Figure \ref{fig_detection_tunning_performance}, the ensemble
improves the performance by ~10\% f1-score over the base model. The ensemble is
composed of a set of tuned models that already outperform the base model. In
the attribution task, the ensemble improves the f1-score by ~9\%, as shown in
Figure \ref{fig_attribution_tunning_performance}.


\subsection{Portability} 
\label{sec_evaluation_portablity}

In this section, we question the portability of the MalDy by applying the
framework on a new type of behavioral reports (Section
\ref{sec_eval_win32_malware_attribution}). Also, we investigate the appropriate
train-set size for MalDy to achieve a good results (Section
\ref{sec_eval_maldy_train_size}). We reports only the results the attribution
task on Win32 malware because we lack Win32 a benign behavioral reports
dataset.

\subsubsection{MalDy on Win32 Malware}
\label{sec_eval_win32_malware_attribution}

Table \ref{tab_viper_attribution_result} presents MalDy attribution performance
in terms of f1-score. In contrast with previous results, we trains MalDy models
on only 2k (10\%)  out of 20k reports' dataset (Table \ref{tab_dataset_list}).
The rest of the reports have been used for testing (18k reports, or 80\%).
Despite that, MalDy achieved high scores that reaches 95\%. The results in
Table \ref{tab_viper_attribution_result} illustrate the portability of MalDy
which increases the utility of our framework across the different platforms and
environments.  

\begin{table}[h!]
\centering
\begin{tabular}{lcc}
\hline
\hline
\textbf{Model}          & \multicolumn{2}{c}{\textbf{Ensemble F1-Score(\%)}}\\
\hline
			& Hashing  	        & TFIDF			 \\
\hline
\hline
CART			& 82.35    		& 82.74 		\\   
ETrees 			& 92.62    		& 92.67 		\\ 
KNN 			& 76.48    		& 80.90 		\\ 
RForest 		& 91.90    		& 92.74 		\\ 
SVM 			& 91.97    		& 91.26 		\\ 
\textbf{XGBoost}	& \textbf{94.86}   	& \textbf{95.43} 	\\ 
\hline
\hline
\end{tabular}
\caption{MalDy Performance on Win32 Malware Behavioral Report}	
\label{tab_viper_attribution_result}
\end{table}                        	

\subsubsection{MalDy Train Dataset Size} 
\label{sec_eval_maldy_train_size}

\begin{figure}[H]
\begin{center}
\subfigure[\scriptsize Hashing (F1-Score \%)]{%
    \label{fig_maldy_viper_trainsize_hashing_performance}
    \includegraphics[width=.5\textwidth]
            {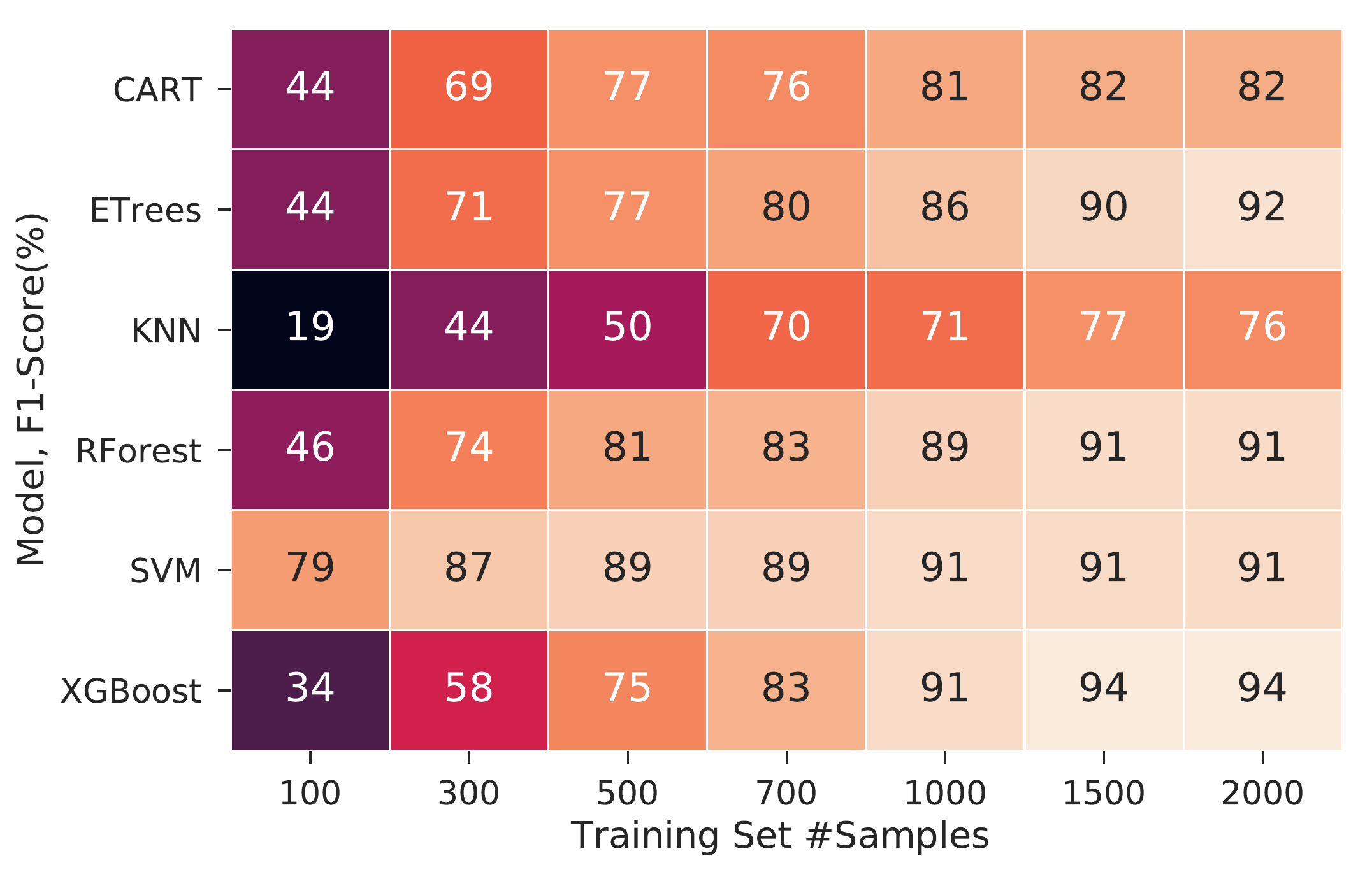}
}\\%
\subfigure[\scriptsize TFIDF (F1-Score \%)]{%
    \label{fig_maldy_viper_trainsize_tfidf_performance}
    \includegraphics[width=.5\textwidth]
            {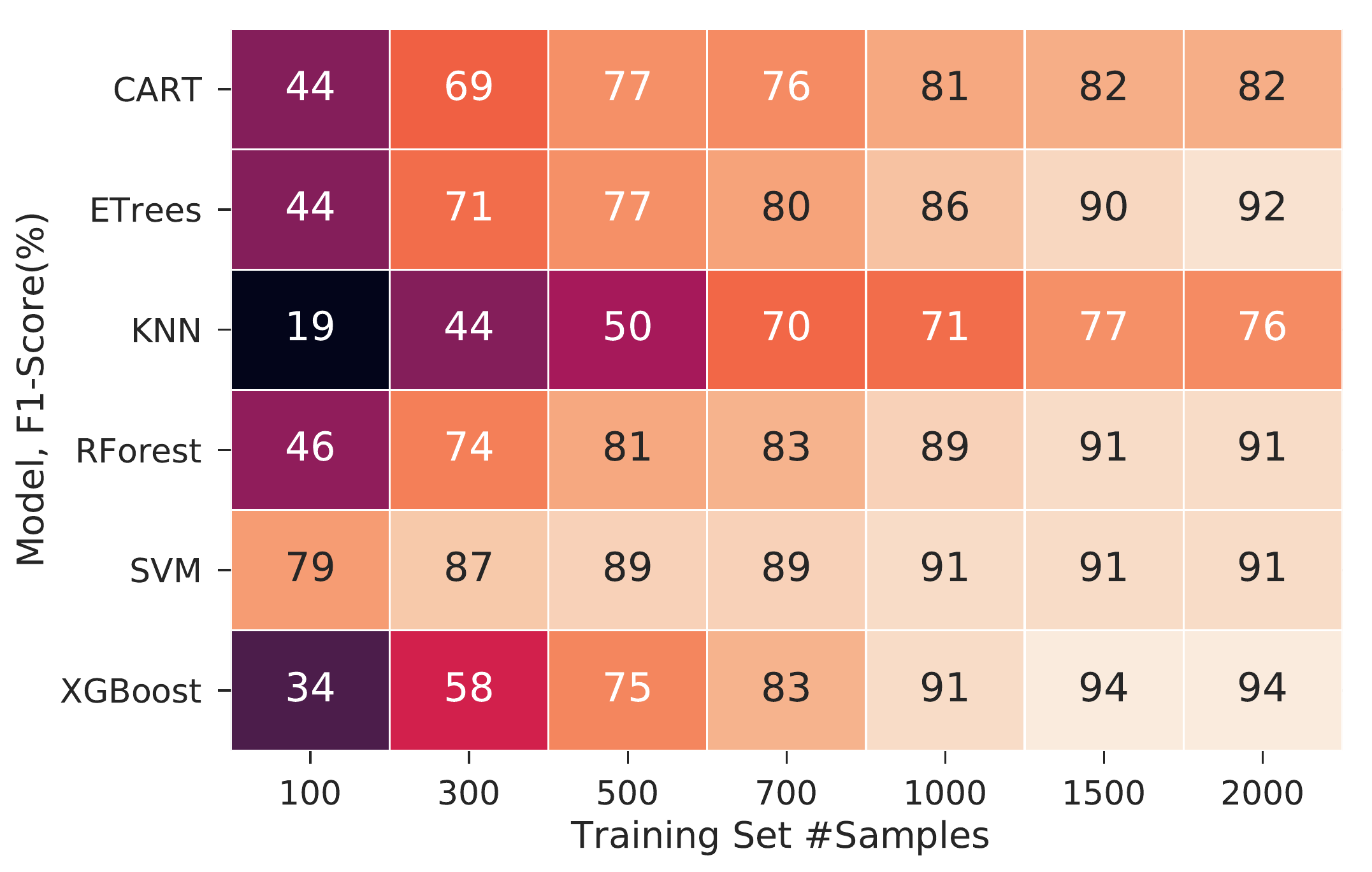}
}%
\end{center}
\caption{MalDy on Win32 Malware and Effect 
            the training Size of the Performance}
\label{fig_maldy_win32_trainsize_effect}
\end{figure}

Using Win32 malware dataset (Table \ref{tab_viper_attribution_result}), we
investigate the train-set size hyper-parameter for Maldy to achieve good
results. Figure \ref{fig_maldy_win32_trainsize_effect} exhibits the outcome of
our analysis for both vectorization techniques and the different classifiers.
We notice the high scores of MalDy even with relatively small datasets. The
latter is very clear if MalDy uses SVM ensemble, in which it achieved 87\%
f1-score with only 200 training samples.

\subsection{Efficiency} 
\label{sec_evaluation_efficiency}

Figure \ref{sec_evaluation_efficiency} illustrates the efficiency of MalDy by
showing the average runtime require to investigate a behavioral report. The
runtime is composed of the preprocessing time and the prediction time. As
depicted in Figure \ref{sec_evaluation_efficiency}, MalDy needs only about 0.03
second for a given report for all the ensembles and the preprocessing settings
except for SVM ensemble. The latter requires from 0.2 to 0.5 seconds (depend on
the preprocessing technique) to decide about a given report. Although SVM
ensemble needs a small train-set to achieve good results (see Section
\ref{sec_eval_maldy_train_size}), it is very expensive in production in terms
of runtime. Therefore, the security investigator could customize MalDy
framework to suite particular cases priorities. The efficiency experiments have
been conducted on Intel(R) Xeon(R) CPU E5\-2630 (128G RAM), we used only one CPU core.

\begin{figure}[H]
\begin{center}
\end{center}
\centering
\includegraphics[width=0.45\textwidth]{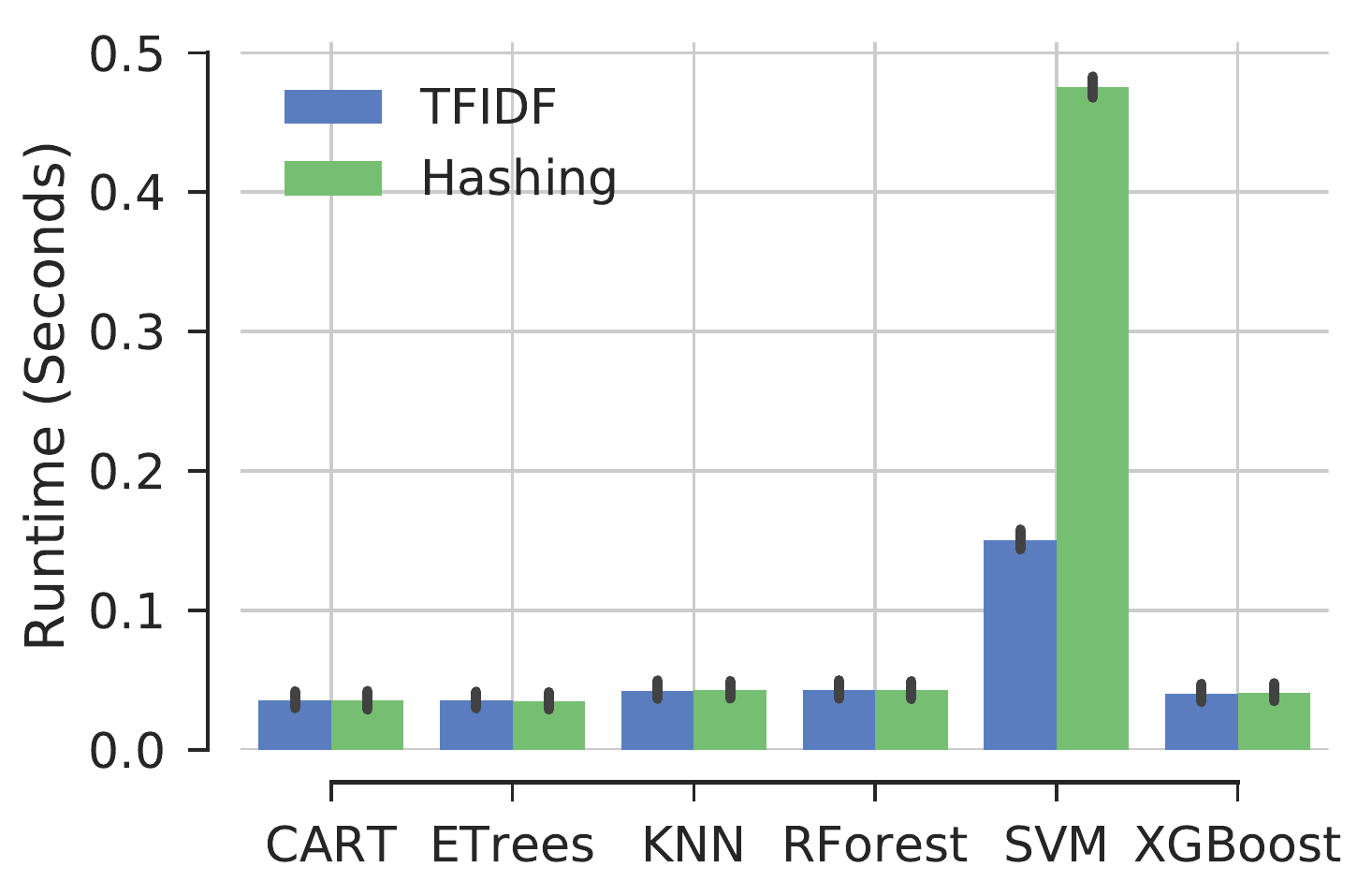}
\caption{MalDy Efficiency}
\label{fig_maldy_efficiency}
\end{figure}

\section{Conclusion, Limitation, and Future Work}

The daily number of malware, that target the well-being of the cyberspace, is
increasing exponentially, which overwhelms the security investigators.
Furthermore, the diversity of the targeted platforms and architectures
compounds the problem by opening new dimensions to the investigation.
Behavioral analysis is an important investigation tool to analyze the binary
sample and produce behavioral reports. In this work, we propose a portable,
effective, and yet efficient investigation framework for malware detection and
family attribution. The key concept is to model the behavioral reports using
the bag of words model. Afterwards, we leverage advanced NLP and ML techniques
to build discriminative machine learning ensembles. MalDy achieves over 94\%
f1-score in Android detection task on Malgenome, Drebin, and MalDozer datasets
and more than 90\% in the attribution task. We prove MalDy portability by
applying the framework on Win32 malware reports where the framework achieved
94\% on the attribution task. MalDy performance depends to the execution
environment reporting system and the quality of the reporting affects its
performance. In the current design, MalDy is not able to measure this quality
to help the investigator choosing the optimum execution environment. We
consider solving this issue for future research.

\section{References}

\bibliographystyle{plain}
\bibliography{reference}

\end{document}